%% file: NonlinearGluing.tex
\documentclass[a4paper,11pt]{article}

\input{packages}

\input{LambdaMacros}

\newcommand{\pt}[1]{\phantom{#1}}
\usepackage{cancel}

\input{WanMacrosKeep.tex}

\input{FinnMacros.tex}
\title{\boldmath Characteristic Gluing with $\Lambda$: III. High-differentiability nonlinear gluing%
\protect\footnote{Preprint: UWThPh 2024-13}
}

 \author[a,b]{Piotr T.\ Chru\'sciel,}
 \author[a,b]{Wan Cong,}
 \author[a]{and Finnian Gray}
 \affiliation[a]{University of Vienna, Faculty of Physics,
 Boltzmanngasse 5, A 1090 Vienna, Austria}

\affiliation[b]{Beijing Institute for Mathematical Sciences and Applications, Huairou, China}

 \emailAdd{wan.cong@univie.ac.at}
 \emailAdd{piotr.chrusciel@univie.ac.at}
 \emailAdd{finnian.gray@univie.ac.at}

\abstract{We prove a nonlinear characteristic $C^k$-gluing theorem for vacuum gravitational fields in Bondi gauge for a class of characteristic hypersurfaces near static vacuum $n$-dimensional backgrounds, $n\ge 3$, with any finite $k$, with cosmological constant $ \Lambda  \in \R$, near Birmingham-Kottler backgrounds.  This generalises  the $C^2$-gluing of Aretakis, Czimek and Rodnianski, carried-out near light cones in four-dimensional Minkowski spacetime.
}

\renewcommand{\blue}[1]{{{#1}}}
\renewcommand{\ptcheck}[1]{}
\renewcommand{\red}[1]{#1}
\renewcommand{\redc}{}

\begin{document}
\maketitle

\input{NonlinearIntroduction}

\input{Acknowledgements}

\input{TheAnalysis}

\input{FunctionSpaces}

\input{BondiSphereDataNewAttempt}

%
%
%
\input{deformations}

\input{NonlinearWanCharge}

\input{argument2a}

\input{NewTheoremv3}
\input{NonlinearProofWan}
\input{argument2}
\newpage
\appendix

\section{ $G_{uA}$ and  $G_{uu}$}
 \label{app18VI23.1}

\input{GuA.tex}

\input{Guu.tex}

\input{charges}
\newpage

\bibliographystyle{JHEP}

\bibliography{NonlinearGluing-minimal}

\end{document}

%% file: packages.tex
\pdfoutput=1 
\usepackage{jheppub} 
\usepackage{makeidx}
\usepackage[totoc]{idxlayout}
\usepackage[T1]{fontenc} 
\definecolor{darkgreen}{rgb}{0.0, 0.5, 0.0}
\usepackage{amsmath}
\usepackage{hhline}
\usepackage{multirow}
\usepackage{rotating}
\usepackage{mathdots}
\usepackage{tikz}
\usetikzlibrary{matrix}
\pgfdeclarelayer{background}
\pgfsetlayers{background,main}
\usepackage{slashed}
\usepackage{url}

\usepackage{amssymb}  
\usepackage{float} 
\usepackage{xcolor} 
\usepackage{graphicx} 
\usepackage{mathdots} 
\usepackage[normalem]{ulem}
\usepackage{bbm}
\usepackage{dsfont}
\usepackage[mathscr]{euscript}
\usepackage{ntheorem}
\usepackage{tensor} 
\usepackage{pdfpages} 

%% file: LambdaMacros.tex
\newcommand{\eqrefl}[1]{\eqref{#1}}

\DeclareFontFamily{U}{mathx}{}
\DeclareFontShape{U}{mathx}{m}{n}{<-> mathx10}{}
\DeclareSymbolFont{mathx}{U}{mathx}{m}{n}
\DeclareMathAccent{\widehat}{0}{mathx}{"70}
\DeclareMathAccent{\widecheck}{0}{mathx}{"71}

\newcommand{\NI}{\blue{\mcN_{[r_1,\fr]}}}
\newcommand{\NIold}{\blue{\mcN_{[r_1,r_2]}}}

\newcommand{\mcNI}{\NI}

\newcommand{\oldkg}{\bluek}

\newcommand{\ralphasq}{\blue{\ell^{-2}}}
\newcommand{\ralpha}{\blue{\ell^{-1}}}

\newcommand{\tsecNtwo}{\blue{\tilde{\secN}_2}}

\newcommand{\mcNnew}{\blue{\mcN_{[r_1,\fr]}}}

\newcommand{\restHtwo}{\blue{|_{\secNtwoNew}}}

\newcommand{\ysqr}{\blue{|y|^2}}
\newcommand{\nusqr}{\blue{|\nu|^2}}
\newcommand{\OneMoreMap}{\blue{\Theta}}

\newcommand{\MainMap}{\blue{\Xi}}
\newcommand{\Mainmap}{\MainMap}

\newcommand{\mcO}{\blue{\mycal{O}}}

\newcommand{\mcG}{\blue{\mycal G}}
\newcommand{\GandD}[1]{\blue{\mycal{G}[#1]}}

\newcommand{\EPsiStar}{\blue{E(\Psi^*}}
\newcommand{\notchi}{\blue{\omega}}

\newcommand{\obstr}{\Im{}}

\newcommand{\bluek}{\blue{k}}
\newcommand{\hak}{\blue{\kgamma}}

\newcommand{\fullmeasure}{\blue{d\mu_{\gamma}}}

\newcommand{\zhTBW}{\blue{\gamma}}
\newcommand{\zzhTBW}{\blue{\mathring{\zhTBW}}}

\newcommand{\ringh}{{  \zzhTBW }}%

\newcommand{\zerogamma}{\zzhTBW}
\newcommand{\twogamma}{\blue{ \zhTBW'}}

\newcommand{\zgamma}{\zerogamma}

\newcommand{\yzero}{\blue{\stackrel{(0)}{y}}}

\newcommand{\fourg}{\blue{g}}

\newcommand{\zfourg}{\blue{\mathring{\fourg}}}
\newcommand{\ztwofourg}{\blue{ \fourg'}}

\newcommand{\zerog}{\blue{\mathring{g}}}
\newcommand{\twog}{\blue{ g'}}

\newcommand{\mzero}{\blue{\mathring{m}}}
\newcommand{\mtwo}{\blue{m'}}

\newcommand{\nobarzg}{\blue{\mathring{g}}}
\newcommand{\zguu}{\blue{\zerog_{uu}}}
\newcommand{\ztwoguu}{\blue{\twog_{uu}}}

\newcommand{\kgamma}{\blue{k_\gamma}}

\newcommand{\cH}{\blue{\widecheck{\mcH}}}

\newcommand{\psizero}{\blue{\psi_0}}

\newcommand{\XN}[1]{\blue{X^{\NI}_{#1} }}
\newcommand{\XNn}[1]{\blue{X^{\mcN}_{#1}}}

\newcommand{\XS}[1]{\blue{X^{\secN}_{#1}}}

\newcommand{\fr}{\blue{\mathring{r}}}

\newcommand{\os}[2]{\blue{\overset{\tiny (#2)}{#1}}}

\newcommand{\cA}{{\check{A}}}
\newcommand{\cB}{{\check{B}}}

\newcommand{\cu}{\blue{\check{u}}}
\newcommand{\cx}{\blue{\check{x}}}
\newcommand{\checkr}{\blue{\check{r}}}

\newcommand{\Psionestar}{}
\newcommand{\Psitwostar}{}

\newcommand{\mepsilon}{\blue{\eta}}

\newcommand{\zsecN}{\blue{\check{\secN}_2}}

\newcommand{\mtdx}{\blue{\check x}}

\newcommand{\mtdu}{\blue{\check u}}
\newcommand{\mtdr}{\blue{\check r}}
\newcommand{\mtdC}{\blue{\check C}}
\newcommand{\mtdD}{\blue{\check D}}

\newcommand{\chx}{\mtdx}

\newcommand{\chu}{\mtdu}
\newcommand{\chr}{\mtdr}

\newcommand{\chC}{\mtdC}

\newcommand{\nur}{\blue{\nu_r}}

{\catcode `\@=11 \global\let\AddToReset=\@addtoreset}
\AddToReset{equation}{section}

{\catcode `\@=11 \global\let\AddToReset=\@addtoreset}
\AddToReset{figure}{section}
\renewcommand{\thefigure}{\thesection.\arabic{figure}}

{\catcode `\@=11 \global\let\AddToReset=\@addtoreset}
\AddToReset{table}{section}
\renewcommand{\thetable}{\thesection.\arabic{table}}

\newcommand{\mcU}{{\mycal U}}

\newcommand{\mcF}{{\blue{\mycal F}}}

\newcommand{\CSdataBo}[1]{{\blue{\Psi_{\mathrm{Bo}}[#1]}}{}}

\newcommand{\CSdata}[1]{{\blue{\Psi[#1]}}{}}

\newcommand{\CdataBo}[1]{{\blue{\Phi_{\mathrm{Bo}}[#1]}}{}}

\newcommand{\fourgtwo}{\blue{\mathbf{g}_2}}

\newcommand{\Czk}{\blue{C^{\bluek}_u\, C^\infty_{(r,x^A)}}}

\newcommand{\Ckr}{\blue{C^{\bluek}_u\, C^\infty_{(r,x^A)}}}
\newcommand{\Ctwo}{\blue{C^2_u\, C^\infty_{(r,x^A)}}}

\newcommand{\genus}{{\blue{\mathfrak{g}}}}

{\catcode `\@=11 \global\let\AddToReset=\@addtoreset}
\AddToReset{figure}{section}
\renewcommand{\thefigure}{\thesection.\arabic{figure}}

{\catcode `\@=11 \global\let\AddToReset=\@addtoreset}
\AddToReset{table}{section}
\renewcommand{\thetable}{\thesection.\arabic{table}}

\newcommand{\harm}{{\blue{\text{H}}}}

\newcommand{\myGauss}{{\blue{\twoscsign}}}

\newcommand{\TTt}{{\blue{\mathrm{TT}}}}

\newcommand{\TS}{\mathop{\mathrm{TS}}}

\newcommand{\twoscsign}{{\blue{\varepsilon}}}

\newcommand{\overadd}[2]
{\overset{ (#1) }{#2}}

\newcommand{\secN}{\blue{\mathbf{S}}}
\newcommand{\secNone}{\blue{\mathbf{S}_1}}
\newcommand{\secNtwo}{\blue{\mathbf{S}_2}}
\newcommand{\mcMtwo}{\blue{\mcM}_2}
\newcommand{\secNtwoNew}{\blue{\check{\mathbf{S}}{}_2}}

\newcommand{\red}[1]{{\color{red} #1}}

\newcommand{\R}{\mathbb{R}}

\newcommand{\mcM}{{\mycal M}}

\newcommand{\mcN}{{\mycal N}}

\newcommand{\sectionofScri}%
{{ \,\,\,\,\mathring{\!\!\!\!\mcN}}}

\newcommand{\redc}{\color{red}}

\newcommand{\eeal}[1]{\label{#1}\end{eqnarray}}

\newtheorem{theorem}{\sc  Theorem\rm}[section]

\newtheorem{Theorem}[theorem]{\sc  Theorem\rm}

\newtheorem{conjecture}[theorem]{\sc  Conjecture\rm}

\newtheorem{definition}[theorem]{\sc  Definition\rm}
\newtheorem{lemma}[theorem]{\sc Lemma\rm}

\newtheorem{proposition}[theorem]{\sc Proposition\rm}

\theorembodyfont{}

\newtheorem{remark}[theorem]{\sc Remark\rm}
\newtheorem{Remark}[theorem]{\sc Remark\rm}

\DeclareFontFamily{OT1}{rsfs}{}
\DeclareFontShape{OT1}{rsfs}{m}{n}{ <-7> rsfs5 <7-10> rsfs7 <10-> rsfs10}{}
\DeclareMathAlphabet{\mycal}{OT1}{rsfs}{m}{n}

\newcommand{\Z}{\mathbbm{Z}}

\newcommand{\blue}[1]{{\color{blue} #1}}

\definecolor{applegreen}{rgb}{0.55, 0.71, 0.0}
\definecolor{armygreen}{rgb}{0.29, 0.33, 0.13}

\definecolor{caribbeangreen}{rgb}{0.0, 0.8, 0.6}

\newcounter{mnotecount}[section]

\renewcommand{\themnotecount}{\thesection.\arabic{mnotecount}}

\newcommand{\mnotex}[1]
{\protect{\stepcounter{mnotecount}}$^{\mbox{\footnotesize
$
\bullet$\themnotecount}}$ \marginpar{
\raggedright\tiny\em
$\!\!\!\!\!\!\,\bullet$\themnotecount: #1} }

\newcommand{\bel}[1]{\begin{equation}\label{#1}}
\newcommand{\bea}{\begin{eqnarray}}
\newcommand{\bean}{\begin{eqnarray}\nonumber}
\newcommand{\beal}[1]{\begin{eqnarray}\label{#1}}
\newcommand{\eea}{\end{eqnarray}}

\newcommand{\nn}{\nonumber}

\newcommand{\qed}{\hfill $\Box$}
\newcommand{\qedskip}{\hfill $\Box$\medskip}
\newcommand{\proof}{\noindent {\sc Proof:\ }}
\newcommand{\be}{\begin{equation}}
\newcommand{\eeq}{\end{equation}}
\newcommand{\ee}{\end{equation}}
\newcommand{\beqa}{\begin{eqnarray}}
\newcommand{\eeqa}{\end{eqnarray}}
\newcommand{\beqan}{\begin{eqnarray*}}
\newcommand{\eeqan}{\end{eqnarray*}}
\newcommand{\ba}{\begin{array}}
\newcommand{\ea}{\end{array}}

\newcommand{\ptcheck}[1]{{\color{darkgreen}\mnotex{ptc : checked on #1}}}

\input{JHmacros}

\newcommand{\spaceD}{{ D}}
\newcommand{\zspaceD}{ {\mathring D}}

\newcommand{\mcH}{\mycal H}

\newcommand{\tmcN}{\blue{\,\,\,\,\,\widetilde{\!\!\!\!\!\mcN}}}



%% file: JHmacros.tex
\def\beq{\begin{eqnarray}}
\def\eeq{\end{eqnarray}}

\def\a{\alpha}
\def\b{\beta}

\def\be{\begin{equation}}
\def\ee{\end{equation}}
\def\bea{\begin{eqnarray}}
\def\eea{\end{eqnarray}}

%% file: WanMacrosKeep.tex
\newcommand{\T}{\mathbb{T}}
\newcommand{\N}{\mathbb{N}}

\newcommand{\ip}[2]{\langle #1, #2\rangle }

\newcommand{\dd}[2]{\frac{\partial #1}{\partial #2} }

\newcommand{\vphi}[1]{\red{\overset{[#1]}{\varphi}}}

\newcommand{\kQ}[2]{\red{\overset{[#1]}{\red{Q}}_{#2}}}

%% file: NonlinearIntroduction.tex
\section{Introduction}

In a recent series of pioneering papers, Aretakis, Czimek and Rodnianski~\cite{ACR2,ACR3} presented a $C^2$-gluing construction near-Minkowskian characteristic initial data for four-dimensional vacuum Einstein equations.
The construction   connects together two spacetimes using a characteristic initial data surface, ensuring continuity of the initial data and of the first two transverse derivatives. The differentiability properties of the spacetime obtained by evolving the resulting initial data are rather poor, when taking into account the differentiability losses arising in the characteristic Cauchy problem. As a result,  the usefulness of the resulting spacetimes for further constructions or applications is limited.

The purpose of this paper is to show how to carry-out the characteristic gluing with an arbitrary finite number of transverse derivatives. While this does not lead to smooth spacetimes by evolution, one can obtain spacetimes which are of arbitrarily high differentiability class, whether classical or Sobolev-type.

We further carry-out the gluing in any spacetime dimension, and allow any cosmological constant $\Lambda\in\R$.

From the point of view of four-dimensional physics, the key contribution of our work is the proof that characteristic gluing in asymptotically Minkowskian four dimensional spacetimes can be carried-out with an arbitrary number of transverse derivatives. As already mentioned, this resolves the issue of poor differentiability of the spacetimes, and hence of the  spacelike initial data sets obtained from spacetimes   evolved from the characteristic data constructed in~\cite{ACR2,ACR3}.  
But the generalisation to higher dimensions and to arbitrary cosmological constants has interest of its own. 

The heart of the proof is to show that the linearised gluing problem can be solved. This has been done in~\cite{ChCong1,ChCongGray1}. One then needs to setup an implicit function theorem, which turns out to be intricate because of intricate differentiability properties of the fields involved. The aim of this work is to carry this out.

\begin{figure}[h]
  \centering
 \includegraphics[scale=0.25]{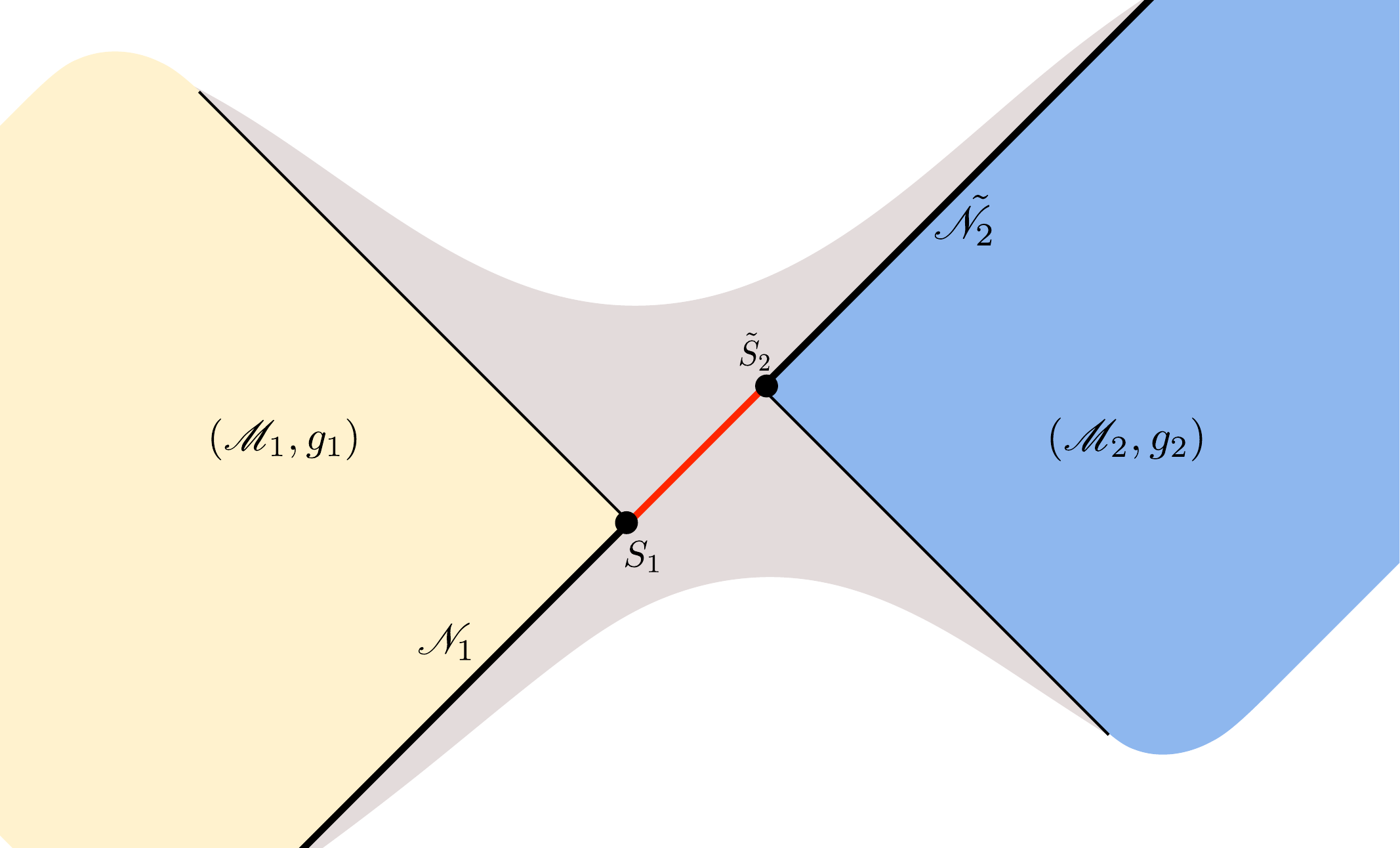}
   \caption{The gluing construction of \cite{ACR2}. Given $\mcN_1\subset \mcM_1$ and $\mcN_2\subset \mcM_2$, the goal is to construct characteristic data interpolating $\mcN_1$ and $\tmcN_2 \subset \mcM_2$, a nearby hypersurface from $\mcN_2$. 
   { The  overlap region between $S_1$ and $\tilde S_2$ is included in $\mcN_1$.}
   Figure adapted from~\cite{ChCong0}.
   }
   \label{F13II23.1}
\end{figure}

To make things precise, the main question of interest is the following:
Consider a
smooth hypersurface $\mcN $ and two characteristic data sets on overlapping subsets $\mcN_1$ and $\mcN_2$ of $\mcN$. Suppose that the data on both $\mcN_1\subset \mcM_1$ and $\mcN_2\subset \mcM_2$ arise by restriction from  vacuum spacetimes  $ (\mcM_1,\fourg_1)$  and $(\mcM_2,\fourg_2)$.  Can one find  a vacuum spacetime $(\mcM ,\fourg)$, with $\mcN\subset \mcM$, so that the data on $\mcN $, arising by restriction from $\fourg$, coincide with the original ones 
\emph{away from the overlapping region}, 
after possibly moving $\mcN_2$ within $\mcM_2$? (compare Figure~\ref{F13II23.1}).

Here we analyse this question for small (nonlinear) perturbations of $(n+1)$-dimensional
Birmingham-Kottler backgrounds, $n\ge 3$;
these include the Minkowski, anti-de Sitter or de Sitter ((A)dS), or Myers-Perry backgrounds.
 In Bondi coordinates the  background metrics can be written as
\be
{\zfourg}\equiv {\nobarzg}_{\a \b} dx^\a dx^\b = \zguu  du^2-2du \, dr
 +
  \underbrace{
   r^2 \ringh_{AB}
   }_{\zerog _{AB}}
   dx^A dx^B
   \,,
   \label{23VII22.3}
\ee
with
$$
\zguu :=
-\big(\twoscsign
 -\ralphasq  r^2- {\frac{2\mzero}{r^{n-2}}}
 \big)
\,,
 \quad
 \twoscsign \in \{0,\pm 1\}
 \,,
 \quad
  \ralpha \in \Big\{0,\sqrt{\frac{2\Lambda}{n(n-1)}}
 \Big \}
  \,,
  \quad
  \mzero \in \R
  \,,
$$
where
$\ringh \equiv \ringh_{AB}dx^A dx^B$ is a $u$- and $r$-independent
Einstein metric of scalar curvature $R[\ringh]$ equal to $(n-1)(n-2)\twoscsign$
 on an $(n-1)$-dimensional manifold $\secN$, which we assume to be compact and boundaryless,
 with the associated
Ricci tensor $R[\ringh]_{AB}$ taking the form
\begin{equation}\label{17III23.1}
  R[\ringh]_{AB} = (n-2) \myGauss\,\ringh_{AB}
  \,,
  \qquad
   \myGauss \in \{0,\pm1\}
   \,.
\end{equation}
Further,  $\ralpha \in \R^+\cup i \R^+$, with a purely imaginary value of $\ralpha$ allowed to accommodate for a cosmological constant $\Lambda <0$.  Finally, the parameter {{$\mzero$}} is related to the total mass of the spacetime.

The hypersurface $\mcN$ will be taken to be $\{u=0\}$, with
$$
\mbox{
$\mcN_1=\{r<r_2\}\cap \mcN$, and  $\mcN_2=\{r>r_1\}\cap \mcN$,}
$$
for some $r_2>r_1>0$.

We will follow the original strategy of \cite{ACR2}, where an implicit function theorem is first used in a form which leads to obstructions to gluing (compare~\cite[Appendix~C]{ChCong0}; see~\cite{CzimekRodnianski} for an alternative approach). Both in~\cite{ACR2} and here one then gets around this problem by considering instead a family of data on a deformation of $\mcN_2$ which carries enough global charges to compensate for these obstructions. In order to account for the obstructions, we will say that a family
$\mcF$  of  \emph{smooth} metrics
defined near $\mcN_2$
is a \emph{compensating family} if $\mcF$  is parameterised diffeomorphically by a set of radial charges obstructing the gluing.
An example in four spacetime-dimensions and with $\Lambda=0$  is provided by the family of boosted-and-translated Kerr metrics.

Let $1\le \bluek \in \N$ be the number of derivatives transverse to $\mcN$ that we want to glue;
the case $k=0$ can be achieved by any smooth interpolation of the unconstrained Cauchy data and does not deserve further considerations.
For simplicity let us at this stage assume that  all fields on $\mcN$ are smooth; this will have to be relaxed in the proof. The space of smooth fields on $\mcN$ with $\bluek$ smooth transverse derivatives will be denoted by $\Ckr$.
As explained in~\cite{ChCong0} the problem of $\Ckr$-gluing of $\mcN_1$ with a deformation of $\mcN_2$ can be reduced to the following:
Let $a\in\{1,2\}$ and
$$\secN_a:= \{u=0\,,\ r=r_a\}
\,,
$$
and let  $x_a \in \CSdata{\secN_a,\bluek }$ be
smooth vacuum codimension-two  data of order $\bluek$
(see Section~\ref{s24VIII23.1}
for the definition)
induced on $\secN_a$ by the codimension-one data on $\mcN_a$.
 One then wants to  find a   vacuum characteristic data set on $\mcN\cap \{r_1\le r\le r_2\}$  which interpolates between $x_1$ and a deformation of $x_2$.

In view of the already-mentioned works on the subject, it is rather clear that the following should be true:

\begin{conjecture}
\label{C5VI23.2}
 Let $
  \bluek
 \in \N$
and let $\mcF$ be a compensating family of
smooth metrics defined near $\mcN_2$.
A smooth, spacelike, vacuum, codimension-two  data set
 $x_1 \in \CSdata{\secN_1,\bluek }$,
   which is sufficiently close in a suitable topology
to the data arising from a member of $\mcF$,   can be $\Czk$-glued
 to data induced on a deformation of  $\secNtwo$ within a nearby
 member of $\mcF$.
\end{conjecture}

In this paper we prove some special cases thereof. The following  is a succinct version of Theorem~\ref{T6V24.1v2} below when $\hak=\infty$:

\begin{Theorem}
 \label{T24VIII23.1}
The conjecture is true near  $(n+1)$-dimensional Birmingham-Kottler metrics, $n\ge 3$, with mass parameter $m\ne 0$,
where $\secN_1$ is a section of the hypersurface $\{u=0\}$ in the coordinate system of \eqref{23VII22.3},
and where $\mcF$ is the family of
\input{Family}
\end{Theorem}

\begin{remark}
 \label{R15VI24.1}
We view the Minkowski metric, the Birmignham-Kottler metrics, the Myers-Perry metrics~\cite{Myers:1987rx}, and their $\Lambda$-counterparts~\cite{Gibbons:2004js,Hawking:1998kw}  as members of  the Kerr-(A)dS family.
From the point of view of the linearised analysis in~\cite{ChCong1,ChCongGray1},
the  metrics missing in \eqref{4V24.1} are the Birmingham-Kottler metrics with a) Ricci-flat sections
$(\secN,\zgamma)$, and b) Einstein sections with positive Ricci tensor
distinct from the round sphere or its quotients.
 This is due to the lack, to the best of our knowledge, of
families of such metrics with enough parameters to compensate for the obstructing radial charges (see  \cite{Klemm:1997ea,Klemm:1998kd} for some partial results).
The existence of any suitable such family  near the Birmingham-Kottler metrics would extend without further due  the range of validity of our gluing results.
\qed
\end{remark}

%% file: Family.tex
\begin{equation}\label{4V24.1}
   \left\{
     \begin{array}{ll}
       \mbox{Kerr-(A)dS metrics}, & \hbox{when $\Lambda \in \R$, $\secNone\approx S^{n-1}$ or a quotient thereof;} \\
       \mbox{Birmingham-Kottler metrics}, & \hbox{when $\Lambda\in \R$, $R[\zgamma] <0$.}
     \end{array}
   \right.
\end{equation}

%% file: Acknowledgements.tex
\noindent{\sc Acknowledgements:} PTC is grateful to Lev Kapitanski for bibliographical advice.\\
\noindent{\sc Declarations}

\noindent{\sc Data availability:} All data for this work are contained within this document.

\noindent{\sc Conflict of Interest:} The authors have no competing interests to declare related to this work.\\

%% file: TheAnalysis.tex
\section{Gluing fields}
  \label{s24VIII23.1}

The aim of this section is to provide a description of the interpolating fields.

Recall that  codimension-two data $\CSdata{\secN ,  \bluek   }$ on a submanifold $\secN$ of codimension two are defined in~\cite{ChCong0} as the collection of jets of order $\bluek$ induced on $\secN$ by smooth Lorentzian metrics defined near $\secN$. Throughout this work we will implicitly  assume that the metric induced on $\secN$ is Riemannian, and that the data satisfy  the differential and algebraic relation following from the vacuum Einstein equations.

We use the parameterisation of the metric   of
Bondi et al.\ (cf., e.g., \cite{MaedlerWinicour,ChCong0} and references therein), namely
\begin{eqnarray}
 \fourg &\equiv&
 g_{\alpha\beta}\mathrm{d}x^{\alpha}\mathrm{d}x^{\beta}
 \nonumber
\\
 &  =  &-\frac{V}{r}e^{2\beta} \mathrm{d}u^2-2 e^{2\beta}\mathrm{d}u \mathrm{d}r
   +r^2\zhTBW_{AB}\Big(\mathrm{d}x^A-U^A\mathrm{d}u\Big)\Big(\mathrm{d}x^B-U^B\mathrm{d}u\Big)
    \, ,
     \label{22XII22.1}
\end{eqnarray}
together with the conditions
\begin{equation}\label{26XII22.1a}
  \partial_r ( \det{\zhTBW_{AB}} ) =
  \partial_u (\det{\zhTBW_{AB}})=
   0
    \,.
\end{equation}
The existence of such coordinates for the class of metrics of interest here  follows from, e.g., \cite[Appendix~B]{ChCong0}.

The Bondi parametrisation of the metric allows one to parameterise $\CSdata{\secN_1,  \bluek   }$ in terms of a reduced set of free data which we denote as $\CSdataBo{\secN_1,  \bluek   }$ (see~\cite{ChCong0} or Section~\ref{s2IV24.1} below).  Now, in~\cite{ChCong0} all  fields have been assumed to be smooth for simplicity, but for the purpose of analysis it is awkward to work with such fields, so that it is useful to make explicit  an index $\hak\in\N$ in $\CSdata{\secN_1,  \bluek  ;\hak}$ to characterise the differentiability class of the fields.
A precise definition  of $ \CSdata{\secN_1,  \bluek;\hak   }$   in terms of the Bondi parameterisation $\CSdataBo{\secN_1, \bluek;\hak}$ is given in Definition~\ref{D12V24.1} below.

It is convenient to assume that the codimension-two data
  $\CSdataBo{\secN_1,  \bluek;\hak }$ and  $\CSdataBo{\secN_2,  \bluek;\hak   }$ arise from vacuum metrics $\fourg_1$ and $\fourg_2$, defined near $\secN_1$ and $\secN_2$ respectively,  both in Bondi gauge with \emph{the same determinant normalisation}, i.e.
   \begin{equation}
   \det \big((\fourg_a)_{AB}\big)
   = r^{2(n-1)}
    \det (\ringh_{AB})
    \,,
    \qquad a=1,2.
    \label{13V24.1a}
    \end{equation}
The Bondi gauge involves no loss of generality for expanding null hypersurfaces, as is the case here, and can be realised while preserving the smallness needed in Theorem~\ref{T24VIII23.1} by e.g.~\cite[Appendix~B]{ChCong0}.
 The metrics $\fourg_1$ and $\fourg_2$ will both be assumed to be close to some background metric $\zfourg$, in norms that are made clear in Theorem
 \ref{T6V24.1v2} below.
   For the purpose of Theorem~\ref{T24VIII23.1} the metric $\zfourg $ will be one of the Birmingham-Kottler metrics with $m\ne 0$ and $\fourg_2$ will be a nearby Kerr-(A)dS metric, or a nearby Birmingham-Kottler metric.

We   choose a number $0<\mepsilon< (r_2-r_1)/16$, such that $\fourg_1$ is
defined on $\{u=0\}$ for $r\le r_1+4\mepsilon$, and that $\fourg_2$ is defined in a neighborhood of $\{u=0\}$ for $r\ge r_2-4\mepsilon$.
The gluing to $\fourg_2$ will take place at $r =\fr$,  a section close to $r=r_2$, where
\begin{equation}\label{29VI24.1}
\fr=\fr(u,x^A)
\end{equation}
is a function which depends upon the data being glued.
The gluing procedure below makes use of a   tensor field $g_{AB}dx^Adx^B$ defined on
\begin{equation}\label{28VI24.11}
 \mcNnew :=\{u=0\}\cap \{x^A\in\secN\,,\ r_1\le r \le \fr(u=0,x^A)\}
 \,;
\end{equation}
this field will interpolate between the given $(\fourg_1)_{AB}$ and $(\fourg_2)_{AB}$ on $\mcNnew$. It takes the form
\begin{equation}\label{8VI23.2}
  g_{AB}
   =\notchi
     \Big(
  \underbrace{
    \zfourg _{AB}
     +
    \phi_1\big((\Psionestar \fourg_1)_{AB} -
    \zfourg _{AB}
     \big)
    +
     \phi_2
     \big(\EPsiStar    \fourg_2)_{AB} -
    \zfourg _{AB}
    \big)
    +
     \sum_{  i\in\iota_{\ralpha,m}  }\kappa_i \, \vphi{i}_{AB}
     }_{=: \hat g_{AB}}
     \Big)
     \,,
\end{equation}
with
\begin{align}
\label{26IV24.1}
 \kappa_i:(r_1,r_2)\rightarrow\R
 \,,
 \quad
    i\in \iota_{\ralpha,m} :=
    \{k_{[\ralpha]},k_{[\ralpha]}+ \frac 12 ,k_{[\ralpha]}+1 ,\ldots,k_{\red{[m]}}+4\} \subset \frac 12 \Z
    \,,
\end{align}
where
\begin{align}
    k_{[\ralpha]} := \begin{cases}
        4-n \,, & \ralpha = 0
        \\
        \min(4-n,\tfrac{7-n-2\bluek}{2}) \,, &\ralpha \neq 0
    \end{cases}
    \,,\qquad
    k_{[m]} := \begin{cases}
        \bluek \,, & m = 0
        \\
        \bluek(n-1) \,, & m \neq 0
    \end{cases}
    \,,
\end{align}
and where the summation ranges originate from the analysis of the linearised problem in~\cite{ChCong1,ChCongGray1}.
In addition:

\begin{enumerate}
 \item
      The function $\notchi>0$ is  determined by the remaining fields appearing in \eqref{8VI23.2} by the requirement that the Bondi determinant condition is satisfied by $\fourg_{AB}$:
 \begin{equation}\label{22VIII23.1}
   \det\big(g_{AB}\big)
    = r^{2(n-1)}  \det\big( \ringh_{AB}
    \big)
     \qquad
     \Longleftrightarrow
     \qquad
     \notchi^{2(n-1)} = \frac{ r^{2(n-1)} \det\big( \ringh_{AB}
    \big) } {\det\big(\hat g_{AB}\big)}
   \,;
 \end{equation}
{{recall that $\hat g_{AB}$ has been defined in \eqref{8VI23.2}.}}

  \item The function $\phi_1=\phi_1(r)$ is a smooth function  which is equal to one for $r\in [r_1,r_1+\mepsilon]$ and vanishes for $r\ge r_1 + 2\mepsilon$, see Figure~\ref{F25VIII23.1}.
      Hence for $r\in [r_1,r_1+\mepsilon]$ we have $g_{AB} = \notchi  \, \Psionestar \,(g_1)_{AB}$, which together with \eqref{13V24.1a} implies that $\notchi \equiv 1$ there. It follows that $g_{AB}$ matches smoothly $ (g_1)_{AB}$ at $\secNone$.
 \begin{figure}
     \centering
     \includegraphics[width=.8\textwidth]{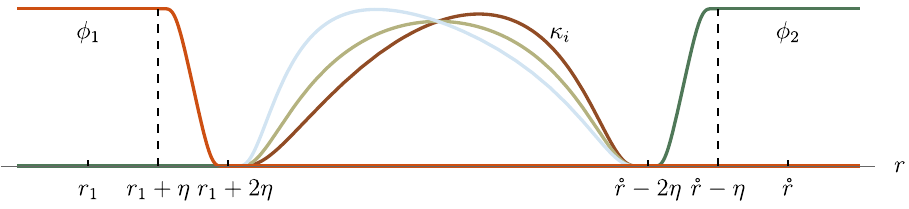}
     \caption{The supports of $\phi_1$, $\phi_2$ and of the $\kappa_i$'s.}
     \label{F25VIII23.1}
 \end{figure}

  \item
  The functions $\kappa_i = \kappa_i(r)$  are smooth,   supported in $[r_1+2\mepsilon, r_2-2\mepsilon]$,
  and satisfy
\begin{eqnarray}
  &
  \displaystyle
 \ip{\kappa_i}{\hat \kappa_j} \equiv \int_{r_1}^{r_2} \kappa_i(s) \hat \kappa_j(s) \, ds = \delta_{ij}
  \,,
  \
 \mbox{where} \ \hat{\kappa}_i(s) :=s^{-i}
  &
  \label{13VIII22.2a}
\end{eqnarray}
(cf., e.g., \cite[Equation~(5.8)]{ChCongGray1}).

  \item The tensor fields $\vphi{i}_{AB}$  are $\zfourg $-traceless, are independent of $r$, and belong to a H\"older space $C^{\hak,\red{\lambda}} (\secN)$ or a Sobolev space $W^{\hak,p}(\secN)$; they are \textit{free fields} which will be used to achieve gluing.

\input{Psi}
  \item
  Let  $0\le \mathring \phi_2:\R\to \R$ be a smooth function  which is equal to one for $x\in [-\mepsilon,\infty)$ and vanishes for $x\le - 2\mepsilon$. We set
  \begin{equation}\label{28VI24.1b}
    \phi_2(r,x^A)= \mathring\phi_2
     \big(r-\fr(u=0,x^A)
    \big)
    \,.
  \end{equation}
      For $r\in [\fr-\mepsilon,\fr]$ we have $g_{AB} = \notchi( \EPsiStar\fourg _2)_{AB})$, which implies that $\notchi \equiv 1$ there. It follows that $g_{AB}$ matches smoothly $\EPsiStar \fourg_2)_{AB}$ at 
      \begin{equation}\label{28VI24.2}
      \tsecNtwo:= \{r=\fr\}\subset \mcNnew
      \,.
      \end{equation}
\end{enumerate}

%% file: Psi.tex
\item $\Psi$ is a diffeomorphism defined in a neighborhood of $\secNtwo$ and preserving the Bondi form of the metric. \textit{The diffeomorphism $\Psi$, together with the fields $\vphi{i}_{AB}$, provides the degrees of freedom needed to achieve gluing.}
The map $E$ is an extension map (see Section \ref{ss6V24/1}), with
 \begin{equation}\label{22VIII23.2}
   \det\big(\EPsiStar \Psitwostar \,\fourg_2)_{AB}\big)
    = r^{2(n-1)}  \det\big( \ringh_{AB}
     \big)
   \,.
 \end{equation}
Note that the right-hand side of \eqref{22VIII23.2}
is chosen once and for all, even if $(\fourg_2)_{AB}$ is another Birmingham-Kottler metric $\ztwofourg_{AB}$.
We emphasise that, in this last case,  the metric $\twogamma_{AB}$ will be different from $\zerogamma_{AB}$ in general, so that \eqref{22VIII23.2} typically requires adjusting the Bondi coordinate $r$; this can be done as follows: Let us write a Birmingham-Kottler metric $\ztwofourg$ 
 as
\be
{\ztwofourg}  = \ztwoguu  du^2-2\,du \, d\rho
 +
   \rho^2 \twogamma_{AB}
   dx^A dx^B
   \,,
   \quad \ztwoguu :=
-\Big(\twoscsign
 -\ralphasq \rho^2- {\frac{2\mtwo}{\rho^{n-2}}}\Big)
\,.
   \label{23VII22.3at2}
\ee
Introducing
\begin{equation}\label{3V24.1138}
  r= \rho\bigg(\frac{\det \twogamma}{\det \zerogamma}
   \bigg)^{\frac{1}{2(n-1)}} =: \rho \chi
\end{equation}
transforms $\ztwofourg$ to a Bondi form
\be
{\ztwofourg} = \ztwoguu  du^2-2\,du \, (\chi^{-1} dr - r\chi^{-2} d\chi)
 +
   r^2 \chi^{-2} \twogamma_{AB}
   dx^A dx^B
\,,
   \label{23VII22.3at21}
\ee
where now \eqref{22VIII23.2} holds: 
 \begin{equation}\label{22VIII23.2345}
   \det\big(\twog_{AB}\big)
    = r^{2(n-1)}  \det\big( \chi^{-2} \twogamma_{AB}
     \big)
    = r^{2(n-1)}  \det\big( \ringh_{AB}
     \big)
   \,.
 \end{equation}

%% file: FunctionSpaces.tex
\section{Definitions, function spaces}
 \label{s2IV24.1}

We need function spaces which are tailored to elliptic equations on $\secN$. As the argument is identical in H\"older spaces and in Sobolev spaces, for $\hak\in \N$, $p\in (1,\infty)$ and $\lambda \in (0,1)$
the space $\XS \hak$ we will use is
\begin{equation}\label{26IV24.21}
  \XS \hak :=
  \left\{
    \begin{array}{ll}
     \mbox{either} \
      C^{\hak,\red{\lambda}}(\secN)\,,
 \\
     \mbox{or} \
      W^{\hak,p}(\secN)
 \,,
    \end{array}
  \right.
\end{equation}
where either the first choice is made throughout, or the second.
The precise values, within the ranges indicated,
 of the Sobolev index $p$ or of the H\"older index $\red{\lambda}$ are irrelevant in the calculations that follow, and are assumed to remain the same throughout the paper. The case $H^\hak\equiv W^{\hak,2}$ is presumably most relevant from the point of view of the evolution problem, but the remaining ones might be of some interest. In what follows, in the Sobolev case the index $\hak$ could in fact be any real number satisfying the inequalities imposed.

Again in the Sobolev case, it might be  useful   to recall the  Moser inequalities on a compact $d$-dimensional manifold $\secN$: for $s\in \R^+$,
 for all
tensor fields   $f\in L^\infty(\secN)$
and  for all smooth  (possibly tensor-valued) maps $F$ there exists a constant
 $C=C(s,F, \|f\|_{L^\infty(\secN)})$ such that
\begin{equation}\label{25VI23.-1}
   \|F(f)\|_{W^{s,p}(\secN)} \le C  \|f\|_{W^{s,p} (\secN)}
   \,.
\end{equation}
We remark that the right-hand side will be finite for $s>d/p$ by Sobolev's embedding, and we will assume throughout that we are in this regime when Sobolev spaces are used.

The manifolds $\NI$ carrying the characteristic data in the gluing region will be of the form  \eqref{28VI24.11}.  We will often simply write $\mcN$ for $\NI$ whenever confusion is unlikely to occur. 

 The following spaces of functions on $\NI$ turn out to be  natural for our problem at hand:
\begin{equation}\label{26IV24.2}
  \XN \hak :=
  \left\{
    \begin{array}{ll}
     \mbox{either} \
\{f
 \  \mbox{such that}
  \
   f \in C^{\hak,\red{\lambda}}(\secNone)\
\mbox{and}
 \
 \partial_r f \in C^{\hak,\red{\lambda}}(\NI)
    \}\,,
 \\
\mbox{or} \
      \{f \ \mbox{such that}\ f\in W^{\hak,p} (\secNone)
      \
 \mbox{and}
 \
 \
 \partial_r f \in  W^{\hak,p} (\NI)
    \}
\,.
    \end{array}
  \right.
\end{equation}

To avoid ambiguities: we will use Sobolev spaces on $\NI$ when boundary data are in Sobolev spaces, and H\"older spaces on $\NI$ when boundary data are in H\"older spaces.

\begin{Remark}
 \label{R25V24.1}
Strictly speaking, the requirement of H\"older regularity of $f$ \emph{in the $r$-direction} is irrelevant for the problem at hand and can be removed from \eqref{26IV24.2}. We use the space $C^{\hak,\red{\lambda}}(\mcN)$  there
to avoid the introduction of yet another nonstandard function space.
\qed
\end{Remark}

\begin{Remark}
 \label{R25V24.1rf} 
Given a $C^1$ function 
{
$\fr=\fr(u,x^A)>r_1$  each of the  maps, parameterised by $u$, defined as
\begin{equation}\label{30VI24.21}
  \NI\ni (r,x^A)
 \mapsto
 \big(
 r_1 + \frac{r_2}{\fr(u,x^A)-r_1}(r-r_1),x^A
 \big)\in \NIold
\end{equation}
}
is a  diffeomorphism. It does \emph{not} preserve the Bondi form of the metric, but for many purposes, e.g.\ for considering the differentiability properties of the fields,  the manifold $\NI$ can be thought of as being the same as $\NIold$, keeping in mind that our functions $\fr(u,\cdot)$ will be $C^1$-close to $r_2$. 
\qedskip
\end{Remark}

We have the following observations, which make it clear how the spaces $\XN \hak$  arise in the calculations below:

\begin{proposition}
  Let $f_\secN\in \XS \hak$ and
$$f_\mcN \in
  \left\{
    \begin{array}{ll}
       C^{\hak,\red{\lambda}}(\NI )\
    &  \mbox{in the H\"older case, or}
 \\  W^{\hak,p} (\NI )
   &  \mbox{in the Sobolev case.}
    \end{array}
  \right.
$$
Then
\begin{enumerate}
  \item At fixed $r$ the functions $ f(r,\cdot) = f_\secN(\cdot)  + \int_{r_1}^{r} f_\mcN(s,\cdot)\,ds  $ are in $ \XS  \hak$, and
  \item The function  $(r,\cdot)\mapsto f_\secN(\cdot) + \int_{r_1}^{r} f_\mcN(s,\cdot)\,ds$ is in $ \XN  \hak$. 
\end{enumerate}
\end{proposition}

\proof
The claims are obvious  in H\"older spaces.

 In $L^p$-type Sobolev spaces with $p\ge 1$, we identify $\NI$ with $\NIold$ as in Remark~\ref{R25V24.1rf}.
We then have
\begin{eqnarray}
  \|\partial_{A_1}\ldots \partial_{A_i} \big(
 f(r,\cdot)-f_\secN(\cdot)
\big)
    \|_{L^p(\secN)}
 &=&
 \big\|
 \int_{r_1}^{r} \partial_{A_1}\ldots \partial_{A_i} f_\mcN(s,\cdot)\,ds
 \big\|_{L^p}
 \nonumber   \\
 &\le &
 \int_{r_1}^{r}  \|\partial_{A_1}\ldots \partial_{A_i} f_\mcN(s,\cdot)
\|_{L^p} ds
 \nonumber   \\
 &\le &
 C(p,r) \Big( \int_{r_1}^{r}  \|\partial_{A_1}\ldots \partial_{A_i} f_\mcN(s,\cdot)
\|^p_{L^p} ds
 \Big)^{1/p}
 \nonumber   \\
 &=  &
 C(p,r) \Big( \int_{r_1}^{r}  \int_{\secN}
|\partial_{A_1}\ldots \partial_{A_i} f_\mcN(s,\cdot)
|^p
 \, d\mu_{\gamma}
 \, ds
 \Big)^{1/p}
 \nonumber   \\
      &=& C(p,r)  \|\partial_{A_1}\ldots \partial_{A_i}  f_\mcN
    \|_{L^p(\mcN_{[r_1,r]})}
 \,,
\end{eqnarray}
{{thus $f(r,\cdot)\in \XS  \hak$. 

Since $f(r_1,\cdot) = f_{\secN}(\cdot) \in W^{\hak,p}(\secN_1)$, to conclude $f \in \XN  \hak$ it remains  to show that $\partial_r f \in  W^{\hak,p} (\NI)$  
  (cf.~\eqref{26IV24.2}). But $\partial_r f = f_{\mathcal{N}}$, which is in $W^{\hak,p} (\NI)$ by hypothesis.}
}
\qedskip

\input{Definitions}

%% file: Definitions.tex
\begin{definition}
 \label{D12V24.1}
 1.
We define spacelike, vacuum, codimension-two Bondi  data   $\CSdataBo{\secN , \bluek;\hak}$ of order $\bluek$, with regularity index $\hak$,   as the following collection of fields on an $(n-1)$-dimensional manifold $\secN $:
\begin{align}
 &
\quad
    V
    \in
 \XS{\hak-2}
  \,,
  \quad
  \partial_rU^A
    \in
 \XS{\hak-1}
    \,,
 \nonumber
 \\
\forall \ 1\le
 \ell \le  \bluek :
 &
\quad
 \partial_r^\ell \gamma_{AB}
    \in
 \XS{\hak+1-\ell}
  \,,
 \nonumber
 \\
\forall \ 0\le
 \ell \le  \bluek :
 &
\quad
 \partial_u^\ell \beta
 \in
 \XS{\hak-2\ell}
 \,,
 \quad
 \partial_u^\ell U^A
     \in
    \XS{\hak-1-2\ell}
    \,,
    \quad
 \partial_u^\ell \gamma_{AB}
     \in
    \XS{\hak-2\ell}
 \,,
 \label{18VI23.11}
\end{align}
where $\gamma_{AB}$ is a Riemannian metric on $\secN$.

2.
We define vacuum, characteristic Bondi  data   $\CdataBo{\mcNI , \bluek;\hak}$ of order $\bluek$, with regularity index $\hak$,   as the following collection of fields on an $n$-dimensional manifold $\mcNI\approx [r_1,r_2]\times \secN$ and on $\secNone:= \{r=r_1\}\subset \mcNI$:
\begin{align}
&
\quad
 \gamma_{AB}
    \in
 \XN{\hak}
  \,,
 \nonumber
 \\
 &
\quad
    V|_\secNone
    \in
 \XS{\hak-2}
  \,,
  \quad
  \partial_rU^A|_{\secNone}
    \in
 \XS{\hak-1}
    \,,
 \nonumber
 \\
\forall \ 0\le
 \ell \le  \bluek :
 &
\quad
 \partial_u^\ell \beta|_\secNone
 \in
 \XS{\hak-2\ell}
 \,,
 \quad
 \partial_u^\ell U^A|_\secNone
     \in
    \XS{\hak-1-2\ell}
    \,,
    \quad
 \partial_u^\ell \gamma_{AB}|_\secNone
     \in
    \XS{\hak-2\ell}
 \,,
 \label{18VI23.11c3}
\end{align}
where each $\gamma_{AB}(r,\cdot)$ is a Riemannian metric on the level sets of $r$ within $\mcNI$.

3. We say that  $\CSdataBo{\secN_1 , \bluek;\hak}$ are compatible with  $\CdataBo{\mcNI , \bluek;\hak}$ if the data induced by the latter at $r=r_1$ coincide with the former.

4. We define a set of  ``deformation-and-gauge fields'' $\GandD{\secN , \bluek;\hak}$ of order $\bluek$, with regularity index $\hak$,   as the following collection of scalars $\psi_i$ and vector fields $\os{X}{i}{}^A$ on  $\secN $:
\begin{align}\label{5IV24.5aagain}
  &
   \psizero
    \in \XS{\hak  +2}
   \,,
   \quad
    \psi_1
    \in \XS{\hak  }
    \,,
    \quad
    \ldots
    \,,
    \quad
  \psi_k
    \in \XS{\hak  +2-2\bluek }
  \,,
\\
 \label{28IV24.1again}
  &
  \os{X}{0}{}^A   \in \XS{\hak  +1}
   \,, \quad
 \os{X}{1}{}^A  \in \XS{\hak  -1}
   \,, \quad
   \ldots
   \,,
   \quad
 \os{X}{\bluek {}}{}^A
   \in \XS{\hak  +1-2\bluek }
  \,.
\end{align}
\qed
\end{definition}

The   fields $\GandD{\secN , \bluek;\hak}$ are used to define the tensor field $\EPsiStar \fourg_2)_{AB}$   in Section~\ref{ss6V24/1}; compare \eqref{5IV24.5aagain2}-\eqref{5IV24.3a4}.

In this terminology,  the set of fields $\CSdataBo{\secN_1, \bluek }$ defined in~\cite{ChCong0} coincides with $\CSdataBo{\secN_1, \bluek;\infty}$; similarly for $\CdataBo{\mcNI , \bluek;\infty}$.

%% file: BondiSphereDataNewAttempt.tex
\section{The equations and their properties}
 \label{s22XII22.2a}

In Definition \ref{D12V24.1}, the information contained in $\CdataBo{\mcNI , \bluek;\hak}$ is equivalent to that contained in  $\CSdataBo{\secN_1 , \bluek;\hak}$ after supplementing it with $\gamma_{AB}\in \XN{\hak}$. The rationale behind the  definition  is, that $\CdataBo{\mcNI , \bluek;\hak}$ allows one to determine the values of the $u$-derivatives of the metric on $\mcN$ up to order $\bluek$:

\begin{Theorem}
 \label{T24VI23.1}
\input{Regularity3v2}%
Let $I\subset \R$ be an interval containing $0$ and let $r_1<\fr:I\times
\secN\mapsto \R$ satisfy 
\begin{equation}\label{28VI24.1}
  0\le i \le \bluek
  \qquad
  \partial^i_u \fr(u, \cdot)   \in \XS{\hak-2i}
  \,.
\end{equation}
The vacuum Einstein equations define a smooth map $\Mainmap$ which to $\fr$ and to the
 characteristic data $\CdataBo{\mcNI , \bluek;\hak}$  satisfying \eqref{18VI23.11c3}
assigns the fields
\begin{align}
   0\le \ell \le  \bluek:  \quad   &
    \partial^\ell _u  \beta
    , \,
     \partial^\ell _u  \gamma_{AB}
 \in \XNn{\hak-2\ell}
 \,,
 \quad
   \partial^\ell _u  U^A ,
   \,
   \partial^\ell _u  \partial_r U^A
 \in \XNn{\hak-1-2\ell}
    \,, \quad
     \partial^\ell _u  V
 \in \XNn{\hak-2-2\ell}
 \,.
 \label{18VI23.10}
\end{align}
%
%
\end{Theorem}

\proof
We can use
Einstein's equations~\cite{ChCong0} (see~\cite{MaedlerWinicour} in spacetime-dimension four)
together with \eqref{25VI23.-1}
to define the following maps:

\begin{enumerate}
 \item We integrate in $r$, 
    within the range $\big[r_1,\fr(u=0,x^A)\big)$, 
    the equation
\begin{equation}
         \label{22XII22.2}
  0 = \frac{r}{2(n-1)} G_{rr} = \partial_{r} \beta - \frac{r}{8(n-1)}\zhTBW^{AC}\zhTBW^{BD} (\partial_{r} \zhTBW_{AB})(\partial_r \zhTBW_{CD})\,.
\end{equation}
This determines
\begin{equation}\label{18VI23.1}
\beta \in \XNn{\hak}
  \,,
\end{equation}
in terms of  $\beta|_{\secNone}\in \XS \hak$ and of the fields on $\secN$. We thus obtain a smooth map
\begin{equation}\label{18VI23.2}
 \Big\{
  \red{\beta|_{\secNone}} \in \XS \hak
     \,,
     \
     \red{\gamma_{AB}\in\XNn{\hak}}
     \Big\}
      \mapsto
     \beta
 \in  \XNn{\hak}
 \,.
\end{equation}
Here (and in what follows), the dependence upon $\fr$ (and its $u$-derivatives) will be kept implicit. 

Assume moreover, for the sake of induction, that there exists $1\le \ell \le \bluek-1$
 such that we have a smooth map  which, to the free data
which are listed in the theorem and which will be made clear as the argument progresses, assigns   the fields
\begin{equation}\label{25VI23.1}
 \mbox{
$
 \partial_u^i   \gamma_{AB}
  \in  \XNn{\hak-2i} $ for $0\le i\le \ell $,
  }
 \end{equation}
 smoothly in the free data.
  Integrating in $r$
    the equation obtained by  differentiating \eqref{22XII22.2} in $u$, we obtain similarly
\begin{equation}\label{18VI23.2i}
 \mbox{
  $
     \partial_u^i  \beta
 \in  \XNn {\hak-2i} $
  for $0\le i\le \ell $,
  }
\end{equation}
smoothly in the free data.

 \item    The fields $U^A|_{\secNone}$ and $\partial_rU^A|_{\secNone}$  are used to obtain $U^A (r,\cdot) $ and $\partial_r U^A (r,\cdot) $ by integrating
\begin{eqnarray}
        0
        &= &
            2r^{n-1}  G_{rA}
             \nonumber
\\
             &= &
              \partial_r \left[r^{n+1} e^{-2\beta}\zhTBW_{AB}(\partial_r U^B)\right]
            -
            2r^{2(n-1)}\partial_r \Big(\frac{1}{r^{n-1}}\spaceD_A\beta  \Big)
                 +r^{n-1}\zhTBW^{EF} \spaceD_E (\partial_r \zhTBW_{AF})
                 \,.
                 \nonumber
                 \\
                            \label{22XII22.3}
           \end{eqnarray}
Combined with \eqref{18VI23.2}, this leads to a smooth map
\begin{align}
 \XS{\hak} \, \oplus
\XS{\hak-1}
 \oplus \XNn{\hak}
  \ni
   &
  \Big(
 \red{\beta|_{\secNone}}\,,
 \,
    \big\{ U^A \,,\, \partial_rU^A
     \big\}|_\secNone,
     \red{\gamma_{AB}}
     \Big)
     \nonumber
 \\
     &
      \mapsto
    \big(   \beta
    \,,\,
     \{U^A   \,,\,
     \partial_r U^A\}
    \big)
 \in  \XNn{\hak}\oplus \XNn{\hak-1}
 \,.
 \label{18VI23.3}
\end{align}
Assuming \eqref{25VI23.1}-\eqref{18VI23.2i}
and
\begin{equation}\label{25VI23.3-}
 \mbox{
  $
     \partial_u^i  U^A|_{\secNone}
     \,,\
      \partial_u^i \partial_r U^A|_{\secNone}
  \in  \XS{\hak-1-2i}$
  for $ 0 \le i\le \ell $,
  }
\end{equation}
by $u$-differentiation
one also finds
\begin{equation}\label{25VI23.3}
 \mbox{
  $
    \partial_u^i  U^A
    ,\,
    \partial_u^i \partial_r  U^A
  \in  \XNn{\hak-1-2i}$
  for $0\le i\le \ell $,
  }
\end{equation}
smoothly in the free data and in $r$.

 \item
  {{Let $R[\gamma]$ denote the Ricci scalar of the metric $\gamma_{AB}$}}. The function $V|_{\secNone}$  is used to integrate in $r$ the equation
 \begin{eqnarray}
         2 \Lambda  r^2
                   &=&
               r^2 e^{-2\beta} (2 G_{ur} + 2 U^A G_{rA} - V/r\, G_{rr} )
               \nonumber
\\
               & = &
                 R[\zhTBW]
                -2\zhTBW^{AB}  \Big[\spaceD_A \spaceD_B \beta
                + (\spaceD_A\beta) (\spaceD_B \beta)\Big]
                +\frac{e^{-2\beta}}{r^{2(n-2)} }\spaceD_A \Big[ \partial_r (r^{2(n-1)}U^A)\Big]
               \nonumber
\\
                &&
                 -\frac{1}{2}r^4 e^{-4\beta}\zhTBW_{AB}(\partial_r U^A)(\partial_r U^B)
                -\frac{(n-1)}{r^{n-3}} e^{-2\beta} \partial_r( r^{n-3} V)
                                 \,,
                  \label{22XII22.4}
           \end{eqnarray}
obtaining thus $V|_{\mcN}$. This results in the smooth map
\begin{align}
 \XS{\hak}
  \,
  \oplus
   &
    \XS{\hak-1}
 \oplus  \XS{\hak-2}
   \oplus \XNn{\hak} \ni
  \Big(
 \red{\beta|_{\secNone}}\,,
 \,
    \big\{ U^A \,,\, \partial_rU^A
     \big\}|_\secNone
    \,,\, V|_\secNone
    \,,
     \
    \red{\gamma_{AB}}
     \Big)
     \nonumber
 \\
     &
      \mapsto
    \Big(  \beta
    \,,\,
  \{ U^A\,,\, \partial_r U^A
  \}
    \,,\,  V
    \Big)
 \in  \XNn{\hak}\oplus  \XNn{\hak-1}\oplus  \XNn{\hak-2}
 \,.
 \label{18VI23.4}
\end{align}
Assuming \eqref{25VI23.1}-\eqref{18VI23.2i}, and \eqref{25VI23.3-} together with
\begin{equation}\label{25VI23.4-}
 \mbox{
  $
     \partial_u^i  V|_{\secNone}
  \in  \XS{\hak-2-2i}$
  for $0 \le i\le \ell $ ,
  }
\end{equation}
one also finds
\begin{equation}\label{25VI23.4}
 \mbox{
  $
      \partial_u^i  V
  \in  \XNn{\hak-2-2i}$
  for $0\le i\le \ell $,
  }
\end{equation}
smoothly in the free data.

 \item     The field $\partial_u \gamma_{AB}|_{\secN}$  is  used to determine  $\partial_u \gamma_{AB}(r,\cdot)$ by integrating
\begin{eqnarray}
 0
 & = &   r^{(n-5)/2} \TS[ G_{AB}]
 \nonumber
\\
& = &
    \partial_r
    \Big[
     r^{(n-1)/2}   \partial_u \zhTBW_{AB}
     	 - \frac{1}{2} r^{(n-3)/2 } V   \partial_r \zhTBW_{AB}
     	 -  \frac{n-1}{4} r^{(n-5)/2  } V    \zhTBW_{AB}
     \Big]
           \nonumber
\\
 && +  \frac{n-1}{4} \partial_r(r^{(n-5)/2  } V )   \zhTBW_{AB}
  \nonumber
  \\
       &&
  +
       \displaystyle
       \frac{1}{2} r^{(n-3)/2}
        V \gamma^{CD} \partial_r\gamma_{AC}\partial_r\gamma_{BD}
       -\frac{1}{2}
        r^{(n-1)/2}
        \gamma^{CD}(\partial_r\gamma_{BD}\partial_u\gamma_{AC}
       +\partial_u\gamma_{BD}\partial_r\gamma_{AC})
       \nonumber \\
       &&
        +
       r^{(n-5)/2}\TS
       \bigg[
      e^{2\beta} r^2  R[\gamma]_{AB}  -2e^{\beta} \spaceD_A \spaceD_B e^\beta
      + r^{3-n} \zhTBW_{CA} \spaceD_B[ \partial_r (r^{n-1}U^C) ]
      \nonumber
\\
 &&
          - \frac{1}{2} r^4 e^{-2\beta}\zhTBW_{AC}\zhTBW_{BD} (\partial_r U^C) (\partial_r U^D)
       +
               \frac{r^2}{2}  (\partial_r \zhTBW_{AB}) (\spaceD_C U^C )
              +r^2 U^C \spaceD_C (\partial_r \zhTBW_{AB})
                \nonumber \\
       &&
        -
	r^2 (\partial_r \zhTBW_{AC}) \zhTBW_{BE} (\spaceD^C U^E -\spaceD^E U^C)
       \bigg]
       \,,
       \label{22XII22.5}
\end{eqnarray}
where the symbol $\TS$ denotes the traceless-symmetric part of a tensor with respect to the metric $\gamma_{AB}$ and where $R[\gamma]_{AB}$ is the Ricci tensor of the metric $\gamma_{AB}$. Hence we obtain the smooth map
\begin{align}
 \XS{\hak} \, \oplus
   &
    \XS{\hak-1}
 \oplus  \XS{\hak-2}
   \oplus \XNn{\hak}
   \ni
   \nonumber
\\
 &
  \Big(
 \red{\beta|_{\secNone}}\,,
 \,
    \big\{ U^A \,,\, \partial_rU^A
     \big\}|_\secNone
    \,,\,
    \big\{
    V\,,\,
    \partial_u \gamma_{AB}
    \}|_\secNone
    \,,
     \
     \red{\gamma_{AB}}
     \Big)
     \nonumber
 \\
     &
      \mapsto
    \Big(  \beta
    \,,\,
     \{ U^A\,,\, \partial_r U^A
  \}
    \,,\,
    \big\{
     V \,,\,
     \partial_u \gamma_{AB}
    \}
    \Big)
 \in  \XNn{\hak}\oplus  \XNn{\hak-1}\oplus  \XNn{\hak-2}
 \,.
 \label{18VI23.5}
\end{align}
Note that this justifies  \eqref{25VI23.1} with $\ell=1$.
Assuming that \eqref{25VI23.1}-\eqref{18VI23.2i}, \eqref{25VI23.3-} and \eqref{25VI23.4-} hold with some $\ell \ge 1$, together with
\begin{equation}\label{25VI23.4bd-}
   \mbox{$
    \partial_u^{\ell+1} \gamma_{AB}|_{\secNone}
  \in  \XS{\hak-2-2\ell -2} $,
  }
\end{equation}
 one also finds
\begin{equation}\label{25VI23.4bd}
 \mbox{$
      \partial_u^{\ell+1}  \gamma_{AB}
  \in  \XNn{\hak-2-2i}$,
  }
\end{equation}
smoothly in the free data and in $r$. Equivalently,  \eqref{25VI23.1} holds with $\ell$ replaced by $\ell+1$.

\item
As already pointed-out, the   $u$-derivative of $\beta$ on $\mcN$ can  be calculated by integrating in $r$ the equation obtained by $u$-differentiating  \eqref{22XII22.2}, after expressing the right-hand side  in terms of the fields determined so far:
\begin{equation}
         \label{22XII22.11}
          \partial_r \partial_u \beta  = \frac{r}{8 }
          \Big(
           \zhTBW^{AC}\partial_u\zhTBW^{BD} (\partial_r \zhTBW_{AB})(\partial_r \zhTBW_{CD})
           +
           \zhTBW^{AC}\zhTBW^{BD} (\partial_r \zhTBW_{AB})(\partial_r\partial_u \zhTBW_{CD})
           \Big)
          \,.
\end{equation}
For this we also need the initial value   $\partial_u  \beta|_{\secN} $, which leads to the smooth map
\begin{align}
 \XS{\hak} \, \oplus
   &
    \XS{\hak-1}
 \oplus  \XS{\hak-2}
   \oplus \XNn{\hak} \ni
    \nonumber
\\
  \Big(
  &
 \red{\beta|_{\secNone}}\,,
 \,
    \big\{ U^A \,,\, \partial_rU^A
     \big\}|_\secNone
    \,,\,
    \big\{
    V\,,\,
    \partial_u \gamma_{AB}
    \,,\,
  \partial_u\beta
    \}|_\secNone
    \,,
     \
     \red{\gamma_{AB}}
     \Big)
     \nonumber
 \\
     &
      \mapsto
    \Big(  \beta
    \,,\,
  \{U^A,\, \partial_r U^A\}
    \,,\,
    \big\{
     V \,,\,
     \partial_u \gamma_{AB}
    \,,\,
     \partial_u \beta
    \}
    \Big)
 \in  \XNn{\hak}\oplus  \XNn{\hak-1}\oplus  \XNn{\hak-2}
 \,.
 \label{18VI23.6}
\end{align}

\item
The equation
\begin{eqnarray}
        -2 e^{2\beta}G_{uA}  =0
\end{eqnarray}
reads
\begin{eqnarray}
0
 &= &
   \partial_r \bigg[
        e^{4\beta}\partial_u\Big(e^{-4\beta}r^2\gamma_{AB}U^B\Big)
      \nonumber
\\
       &  &
      \phantom{
   \partial_r \bigg[}
       -
         e^{2\beta} \partial_r \Big( r\gamma_{AB}U^B V e^{2\beta}\Big) -2 r V \partial_r (\gamma_{AB}U^B)
    + r^2 U^B\partial_u\gamma_{AB}
    \bigg]  +
 \mathcal{F}_A
     \,,
     \phantom{xxx}
      \label{1V22.12}
\end{eqnarray}
%
%
where $\mathcal{F}_A$ can be read-off from
Appendix~\ref{app18VI23.1}.
This equation
allows us to determine algebraically $\partial_r\partial_u U^A (r_1,\cdot)$ in terms of  fields which have been determined in the previous steps.
One thus obtains a   smooth map
\begin{align}
 \XS{\hak}& \oplus
    \XS{\hak-1}
 \oplus  \XS{\hak-2}
 \oplus  \XS{\hak-3}
   \oplus \XNn{\hak}
  \ni
  \nonumber
  \\
  \Big(
  &
 \red{\beta|_{\secNone}}\,,
 \,
    \big\{ U^A \,,\, \partial_rU^A
     \big\}|_\secNone
    \,,\,
    \big\{
    V\,,\,
    \partial_u \gamma_{AB}
    \,,\,
  \partial_u\beta
    \}|_\secNone
    \,,\,
  \partial_uU^A|_\secNone
    \,,
     \
     \red{\gamma_{AB}}
     \Big)
     \nonumber
 \\
     &
      \mapsto
    \Big(  \beta
    \,,\,
  \{U^A,\, \partial_r U^A\}
    \,,\,
    \big\{
     V \,,\,
     \partial_u \gamma_{AB}
    \,,\,
     \partial_u \beta
    \}
    \,,\,
  \{
     \partial_u U^A,\,
     \partial_u \partial_r U^A\}
    \Big)
     \nonumber
 \\
  &
 \in  \XNn{\hak}\oplus  \XNn{\hak-1}\oplus  \XNn{\hak-2}\oplus  \XNn{\hak-3}
 \,.
 \label{18VI23.7}
\end{align}
Note that this shows that the $\partial_u^i \partial_r U^A$-part of \eqref{25VI23.3-}
holds with $\ell=1$.
 We remark that the consistency of this equation with the one obtained by $u$-differentiating \eqref{22XII22.3}  follows from Bianchi identities.

 \item We can determine algebraically  $\partial_uV$ on $\secN$  from the Einstein equation $
     (G_{uu}+\Lambda g_{uu})|_\mcN = 0$:
  \begin{equation}\label{13I23.1}
    G_{uu} = \frac{n-1}{2 r^{2}}\partial_u V + ...
    \,,
  \end{equation}
  where ``$...$'' stands for an explicit expression in all fields already known on $\mcN$, see Appendix~\ref{app18VI23.1}.
This shows that  \eqref{25VI23.4-}
holds with $\ell=1$.

\end{enumerate}

  The whole argument so far leads thus to a smooth map
\begin{align}
 \XS{\hak}& \oplus
    \XS{\hak-1}
 \oplus  \XS{\hak-2}
 \oplus  \XS{\hak-3} \oplus
  \XNn{\hak} \ni
  \nonumber
  \\
  &
  \Big(
 \red{\beta|_{\secNone}}\,,
 \,
    \big\{
      U^A,\,
     \partial_rU^A
     \big\}|_\secNone
    \,,\,
    \big\{
    V\,,\,
    \partial_u \gamma_{AB}
    \,,\,
  \partial_u\beta
    \}|_\secNone
    \,,\,
     \partial_u U^A |_\secNone
    \,,
     \
     \red{\gamma_{AB}}
     \Big)
     \nonumber
 \\
     &
      \mapsto
    \Big(  \beta
    \,,\,
  \{ U^A
 ,\, \partial_r U^A
 \}
    \,,\,
    \big\{
     V \,,\,
     \partial_u \gamma_{AB}
    \,,\,
     \partial_u \beta
    \}
    \,,\,
  \{
     \partial_u U^A,\,
     \partial_u \partial_r U^A\}
    \,,\,
     \partial_u V
    \Big)
     \nonumber
 \\
  &
 \in  \XNn{\hak}\oplus  \XNn{\hak-1}\oplus  \XNn{\hak-2}\oplus  \XNn{\hak-3}
  \oplus \XNn{\hak-4}
 \,.
 \label{18VI23.8}
\end{align}
 (Similarly to \eqref{22XII22.3} and \eqrefl{1V22.12}, the consistency of \eqref{18VI23.8}  with the equation obtained by $u$-differentiating \eqref{22XII22.4}  follows from Bianchi identities.)

One can  inductively repeat the procedure above using the equations obtained by differentiating Einstein equations with respect to $u$. This finishes the proof.
\qed
%

%% file: Regularity3v2.tex
Let $\bluek \in \N$, $\hak\in \N\cup\{\infty\}$.   We suppose that, in $n$-space dimensions, $n \ge 3$, the regularity index  $\hak$ satisfies
\begin{equation}\label{n25V24.11Xp}
 \hak 
  \left\{
    \begin{array}{ll}
     \ge  2+2\bluek
    &  \mbox{in the H\"older case, or}
 \\ > 2+2\bluek +(n-1)/p \
   &  \mbox{in the $L^p$-type Sobolev case.}
    \end{array}
  \right.
\end{equation}

%% file: deformations.tex
%
%
%
%

\input{NonlinearGauge}

\input{OnStwo}
\input{nonzeronuA.tex}

\input{Continuity}

%% file: NonlinearGauge.tex
\section{Deforming $\secNtwo$}
 \label{s26VI23.1}

The aim of this section is to provide a parametrisation of the map $\Psi $  appearing in \eqref{8VI23.2}.
This requires  an analysis of coordinate transformations which preserve  the
null-hypersurface form of the metric
\begin{equation}
    \fourg = -\alpha du^2 + 2\nur  du dr + 2 \nu_A du dx^A + g_{AB} dx^A dx^B
    \,,
     \label{19VII23.11}
\end{equation}
together with the Bondi  determinant-conditions
\begin{equation}\label{21III22.2}
 | \partial_{r}
 (  {\det  g _{AB}}
  )
   |
 >0
   \,,
   \quad
 \partial_{r} \big(
  r^{-2(n-1)} {\det  g _{AB}}
  \big)
   = 0
    = \partial_{u} \big(
 {\det  g _{AB}}
  \big)
 \,.
\end{equation}
Thus, consider  a  coordinate transformation $x^{\mu}\rightarrow \mtdx^{\mu}\equiv (\mtdu,\mtdr,\mtdx^{A})$. It is convenient to
write
(hoping that no confusion will arise with the field $U^A$  of Section~\ref{s22XII22.2a} and the   field $U_{\check{A}}$ here) 
\begin{align}
    \dd{u}{\mtdu} &= U_{\mtdu}\,,\quad \dd{r}{\mtdu} = R_{\mtdu} \,,\quad \dd{u}{\mtdr} = U_{\mtdr} \,,\quad
    \dd{r}{\mtdr} = R_{\mtdr} \,,
\\
    \dd{u}{\mtdx^C} &= U_{\check C}\,,\quad
    \dd{r}{\mtdx^C} = R_{\check C}\,,\quad
    \dd{x^A}{\mtdu} = X^A_{\mtdu}\,,\quad
    \dd{x^A}{\mtdr} = X^A_{\mtdr}\,,\quad
    \dd{x^A}{\mtdx^B} = \red{\Lambda}^A_{\pt{A}\check B}\,.
    \end{align}
It holds that
\begin{eqnarray}
\fourg
   &\to &
 \bigg[ g_{AB} X^A_{\mtdu} X^B_{\mtdu}  -  \alpha U_{\mtdu}^2   +   2 \nur  R_{\mtdu} U_{\mtdu} + 2 \nu_A U_{\mtdu}X^A_{\mtdu}\bigg] d\mtdu^2 \nonumber
\\
& &
+ \bigg[ g_{AB} X^A_{\mtdr} X^B_{\mtdr}  -  \alpha U_{\mtdr}^2  +   2 \nur  R_{\mtdr} U_{\mtdr} + 2 \nu_A U_{\mtdr}X^A_{\mtdr}\bigg] d\mtdr^2 \nonumber
\\
& &
+ \bigg[g_{AB} \red{\Lambda}^A_{\pt{A}\check C} X^B_{\mtdu} - \alpha U_{\mtdu} U_{\check C} + \nur  (R_{\check C} U_{\mtdu} +R_{\mtdu} U_{\check C} )
   +\nu_A ( U_{\check C}X^A_{\mtdu} +U_{\mtdu}\red{\Lambda}^A_{\pt{A}\check C} ) \bigg] 2 d\mtdu \, d\mtdx^C \nonumber
   \\
& &
+ \bigg[g_{AB} \red{\Lambda}^A_{\pt{A}\check C} X^B_{\mtdr} - \alpha U_{\mtdr} U_{\check C} + \nur  (R_{\check C} U_{\mtdr} +R_{\mtdr} U_{\check C} )
   +\nu_A ( U_{\check C}X^A_{\mtdr} +U_{\mtdr}\red{\Lambda}^A_{\pt{A}\check C} ) \bigg] 2 d\mtdr d\mtdx^C \nonumber
\\
& &
+\left[g_{AB} X^A_{\mtdu}X^B_{\mtdr} -\alpha U_{\mtdr}U_{\mtdu} + \nur  (R_{\mtdu} U_{\mtdr} +R_{\mtdr} U_{\mtdu} ) + \nu_A (U_{\mtdu}X^A_{\mtdr} + U_{\mtdr} X^A_{\mtdu} ) \right] \, 2 d\mtdu \, d\mtdr
 \nonumber
\\
& &
+\left[g_{AB} \red{\Lambda}^A_{\pt{A}\check C} \red{\Lambda}^B_{\pt{B}\check D} + U_{\check C} ( 2\nur  R_{\check D} + 2\nu_A  \red{\Lambda}^A_{\pt{A}\check D} -\alpha U_{\check D} )\right] d\mtdx^C d\mtdx^D
 \,.
  \label{20III221b}
\end{eqnarray}
To preserve the null form of the metric  we need,
\begin{align}
     g_{AB} X^A_{\mtdr} X^B_{\mtdr}  -  \alpha U_{\mtdr}^2  +   2 \nur  R_{\mtdr} U_{\mtdr} + 2 \nu_A U_{\mtdr}X^A_{\mtdr} & = 0 \,,
       \label{21III22.1.1a}
\\
    g_{AB} \red{\Lambda}^A_{\pt{A}\check C} X^B_{\mtdr} - \alpha U_{\mtdr} U_{\check C} + \nur  (R_{\check C} U_{\mtdr} +R_{\mtdr} U_{\check C} )
   +\nu_A ( U_{\check C}X^A_{\mtdr} +U_{\mtdr}\red{\Lambda}^A_{\pt{A}\check C} ) &= 0
 \,,
   \label{21III22.1again}
\end{align}
while   Bondi coordinates require in addition
the $\mtdx^\mu$-equivalent of the determinant condition \eqref{21III22.2}.

%% file: OnStwo.tex
We concentrate first on \eqref{19VII23.11} and its hatted equivalent, ignoring momentarily  both  \eqref{21III22.2} and its hatted equivalent. In the first two steps of our construction we will
restrict ourselves to coordinate transformations for which
\begin{equation}\label{12IX23.1}
  \mtdr \equiv r
  \,,
\end{equation}
so that
\begin{align}
    \dd{r}{\mtdr} =
    \dd{\mtdr}{r}=
    1
    \,,
    \quad
    \dd{r}{\mtdu} =
    \dd{r}{\mtdx^C} =
    \dd{\mtdr}{u} =
    \dd{\mtdr}{x^A} = 0\,.
    \end{align}

The equations simplify somewhat if $\nu_A=0$. We then have
\begin{eqnarray}
 \fourg
   &= &
 \big[ g_{AB} X^A_{\mtdu} X^B_{\mtdu}  -  \alpha U_{\mtdu}^2
  \big] d\mtdu^2
 \nonumber
\\
& &
 + 2
 \big[g_{AB}
 \red{\Lambda}^A_{\pt{A}\check C} X^B_{\mtdu}
 - \alpha U_{\mtdu} U_{\check C}
 \big]
d\mtdu \, d\mtdx^C
+
 2 \big[g_{AB} X^A_{\mtdu}X^B_{\mtdr} -\alpha U_{\mtdr}U_{\mtdu}
  + \nur    U_{\mtdu}
 \big]
   d\mtdu \, d\mtdr
 \nonumber
\\
& &
 +
\big[
 \underbrace{
  g_{AB} \red{\Lambda}^A_{\pt{A}\check C} \red{\Lambda}^B_{\pt{B}\check D}
-\alpha U_{\check C}  U_{\check D}
    }_{g_{\mtdC\mtdD}}
\big]
d\mtdx^C d\mtdx^D
 \,,
  \label{20III221bcd}
\end{eqnarray}
with
\begin{align}
     g_{AB} X^A_{\mtdr} X^B_{\mtdr}  -  \alpha U_{\mtdr}^2  +   2 \nur  U_{\mtdr}
       & = 0 \,,
       \label{21III22.1.1n}
\\
    g_{AB} \red{\Lambda}^A_{\pt{A}\check C} X^B_{\mtdr} - \alpha U_{\mtdr} U_{\check C}
     + \nur   U_{\check C}
   &= 0
 \,.
   \label{19VII23.21}
\end{align}
Equations
 \eqref{21III22.1.1n}-\eqref{19VII23.21} imply
\begin{align}
    (\alpha U_{\mtdr} - \nur  )^2
    \underbrace{ U_{\check C}{U_{\check D}} (\Lambda^{-1})^{\check C}{}_{A}  (\Lambda^{-1})^{\check D}{}_{ E} g^{EA} }_{=: \ysqr  }
    = (\alpha U_{\mtdr}-2 \nur  )U_{\mtdr} \,.
    \label{20VII23.w1}
\end{align}
From now on we assume that
$$
 \alpha\, \ysqr  <1
 \,,
$$
as needed to solve the quadratic equation \eqref{20VII23.w1} for a real-valued function $U_{\mtdr}$.
The relevant solution  is the one which is small when $\ysqr$ is small:
\begin{equation}\label{18VIII23.f1}
    U_{\mtdr}
    =
   -\frac{\nur  \ysqr  }{ 1-\alpha \ysqr  +(1-\alpha \ysqr )^{1/2}  } \,.
\end{equation}
This allows us to rewrite  \eqref{19VII23.21} as
\begin{equation}\label{18VIII23.f2}
    X^A _{\mtdr}
    =
   -  \frac{\nur   }{   (1-\alpha \ysqr )^{1/2}  }
  \underbrace{g^{AB} (\Lambda^{-1})^{ \chC}{}_B{}  U_{\chC}}_{=:y^A}\,.
\end{equation}
Inserting \eqref{18VIII23.f1}-\eqref{18VIII23.f2}  into \eqref{20III221bcd} we obtain the following $g_{\mtdr\mtdu}$-component of the metric:
\begin{equation}
	g_{\mtdr\mtdu}=\frac{\nur   }{   (1-\alpha \ysqr )^{1/2}  }
        \big(U_{\mtdu}- U_{\chC} (\Lambda^{-1})^{ \chC}{}_A{} X^A_{\mtdu}
         \big) \,.
         \label{11IV24.13}
\end{equation}

%% file: nonzeronuA.tex
When $\nu_A$ is nonzero, as is the case in \eqref{23VII22.3at21}, we have to solve the full equations \eqref{21III22.1.1a}-\eqref{21III22.1again} for $ X^A_{\mtdr}$ and $U_{\mtdr}$. We continue to assume that $\mtdr \equiv r$.  
It is convenient to define the   fields 
\begin{align}
\label{7V24.21}
	\theta_\mtdr
 : =\nu_AX^A_{\mtdr}\,,\quad Y^{A}_{\mtdr}
 :=X^A_{\mtdr}-\theta_\mtdr\frac{\nu^A}{\nusqr }\,,\quad
\end{align}
{{where
\begin{equation}
 \nu^A = g^{AB}\nu_B \,, \quad \nusqr =g^{AB}\nu_A\nu_B
 \,.
\end{equation}
}}
Equations~\eqref{21III22.1.1a}-\eqref{21III22.1again} 
expressed in terms of  these fields become
\begin{align}
g_{AB} \left(Y^A_{\mtdr}+\theta_\mtdr\frac{\nu^A}{\nusqr }\right) \left(Y^B_{\mtdr}+\theta_\mtdr\frac{\nu^B}{\nusqr }\right) &  = ( \alpha U_{\mtdr}  -  2 \nur - 2 \theta_{\mtdr}) U_{\mtdr} \,,
\label{3V24.f2a}
\\
 Y^A_{\mtdr} + (\nusqr  U_{\mtdr}+\theta_{\mtdr})\frac{\nu^A}{\nusqr } &= (\alpha U_{\mtdr}  - \nur -\theta_{\mtdr} ) y^A
\,.
\label{3V24.f2b}
\end{align}
Contracting \eqref{3V24.f2b} with $\nu^A$ gives an expression for $\theta_{\mtdr}$ in terms of $U_{\mtdr}$ and of the metric functions:
\begin{align}\label{3V24.f3}
	\theta_{\mtdr}=\frac{(\alpha \nu_Ay^A-\nusqr )U_{\mtdr}-\nur y^A\nu_A}{1+y^A\nu_A}\,.
\end{align}
Next, we can find another equation relating $\theta_{\mtdr}$ and $U_\mtdr$ by calculating  $g_{AB} Y^A_{\mtdr}Y^B_{\mtdr}$ using \eqref{3V24.f2b}. After this \eqref{3V24.f2a}-\eqref{3V24.f2b} become, using $Y^{A}_{\mtdr}\nu_{A}=0$,
\begin{align}
g_{AB} Y^A_{\mtdr}Y^B_{\mtdr}+\frac{\theta^2_\mtdr}{\nusqr } &  = ( \alpha U_{\mtdr}  -  2 \nur - 2 \theta_{\mtdr}) U_{\mtdr} \,,
\label{3V24.f4a}
\\
g_{AB}Y^A_{\mtdr}Y^B_{\mtdr} + \frac{(\nusqr  U_{\mtdr}+\theta_{\mtdr})^2}{\nusqr } &= (\alpha U_{\mtdr}  - \nur -\theta_{\mtdr} )^2 y^Ay_A\,.
\label{3V24.f4b}
\end{align}
Eliminating $g_{AB} Y^A_{\mtdr}Y^B_{\mtdr}$ yields
\begin{equation}
	( \alpha U_{\mtdr}  -  2 \nur - 2 \theta_{\mtdr}) U_{\mtdr}-\frac{\theta^2_\mtdr}{\nusqr }=(\alpha U_{\mtdr}  - \nur -\theta_{\mtdr} )^2 y^Ay_A- \frac{(\nusqr  U_{\mtdr}+\theta_{\mtdr})^2}{\nusqr }
\end{equation}
which, upon substituting \eqref{3V24.f3}, leads  to the following quadratic equation for $U_\mtdr$:
{{
\begin{equation}
		(\alpha+\nusqr )[(1+y^A\nu_A)^2-\ysqr (\alpha+\nusqr )]U_{\mtdr}^2
 -2[(1+y^A\nu_A)^2-\ysqr (\alpha+\nusqr )] \nur U_\mtdr- \ysqr \nu_r^2=0\,.
  \label{19VI24.1}
\end{equation}
}}
Let us assume that  
\begin{equation}\label{19VI24.2}
 z
  :=\frac{(1+y^A\nu_A)^2}{\alpha+\nusqr }
> \ysqr 
 \,,
\end{equation}
with $\ysqr =g_{AB}y^Ay^B$ as in \eqref{20VII23.w1},
which is clearly true for sufficiently small $y^A$, as needed below.
Then the solutions to \eqref{19VI24.1} are real-valued and equal to 
\begin{eqnarray}
	U_\mtdr & 
 =
  &
  \frac{ \left(z-\ysqr \pm \sqrt{z \left(z-\ysqr \right)}\right)}{\left(\alpha +\nusqr\right) \left(z-\ysqr \right)}\nur
    =:  F(\partial_{\cA} u, \partial_{\cB} x^{A})\,.
     \label{3V24.p0} 
\end{eqnarray}
We take the negative root, which reduces to \eqref{18VIII23.f1} in the limit $\nu_A\to0$.

Finally, we can write down our solution for $X^A_\mtdr$ substituting \eqref{3V24.f3} into \eqref{3V24.f2b} and recalling $Y^{A}_{\mtdr}=X^A_{\mtdr}-\theta_\mtdr\nu^{-2}\nu^A$. This gives,
\begin{align}
X^A_{\mtdr}=& - U_\mtdr \nu^A + \left(\frac{(\alpha+\nusqr )U_\mtdr-\nur}{1+y^A\nu_A}\right)y^A\nonumber\\
  =&  -\nur\frac{ \left(z-\ysqr  - \sqrt{z \left(z-\ysqr \right)}\right)}{\left(\alpha +\nu ^2\right) \left(z-\ysqr \right)}\nu^A -\frac{\nur }{\sqrt{\left(z-\ysqr \right)(\alpha+\nusqr )}}y^A
\nonumber\\
     =: & \ F^A(\partial_\cA u, \partial_\cB x^ A)
   \,.
    \label{3V24.p1}
\end{align}
The key fact for us is that the functions $F$ and $F^A$ defined in \eqref{3V24.p0}-\eqref{3V24.p1} are smooth functions of their arguments and of the metric coefficients when   the derivatives $\partial_\cA u $ are small.
%

%% file: Continuity.tex
\subsection{Regularity}
 \label{s18VIII23.1}

The gluing construction of~\cite{ACR2,ACR3}  requires a deformation of the section
$$
 \secNtwo= \{r=r_2\}\cap \NIold
$$
in the spacetime $(\mcMtwo,\fourgtwo)$, as well as a prescription for the calculation of the $u$-derivatives of this deformation. This is  needed to control some of the gauge-dependent
radially-conserved charges. One needs furthermore to include in the construction a diffeomorphism $\Phi^A$ of $\secNtwo$, as well as its $u$-derivatives. Last but not least, one  needs to make sure that  the regularity of the resulting fields  is consistent with the characteristic constraint equations and their $u$-derivatives.

Now, our aim is to provide a scheme to which the implicit function theorem can be applied. This puts stringent requirements on the differentiability properties of the fields at hand, and makes the construction demanding. We note that trying to do all the coordinate changes at once, or changing the order of the coordinate transformations below, or introducing $\chu$ as a function of $u$ rather than $u$ as a function of $\chu$, etc., leads to fields with problematic regularity properties.
 
In the original coordinate system the new section, which we denote by $\secNtwoNew$,  will be  given by the equations
\begin{equation}\label{30III24.1}
 \secNtwoNew=
  \{
    u =\psizero(x^A)
   \,,
  \
        r=r_2
    \}
    \subset
  \{
    r=r_2
    \}
    =: \cH_2
   \,,
\end{equation}
with a function  $\psizero$ which will be determined in the course of the proof of Theorem~\ref{T6V24.1v2}. After carrying-out this deformation, 
for the purpose of this last theorem we will need to adjust the coordinates $x^A$ on $\secNtwoNew$, and to adjust the coordinate $r$ on $\NIold$ which will   determine the function $\fr$ of \eqref{29VI24.1}.

\input{RemarkDiff}

So let  $1\le \bluek\in \N\cup\{\infty\}$ be the number of transverse derivatives which we wish to glue. Let $ \hak\in \N\cup\{\infty\}$
satisfy
 \begin{equation}\label{7V24.1}
    \hak/2 - \red{1}
     \,
       \begin{cases}
         \ge\bluek  \,, & \mbox{in the H\"older case,}
         \\
          >\bluek+(n-1)/2p \,, & \mbox{in the $L^p$-type Sobolev case.}
     \end{cases}
 \end{equation}
Recall that  $\hak$  encodes the differentiability properties of the fields.

In the calculations that follow we work on a  spacetime manifold $\mcM$ satisfying 
\begin{equation}\label{26V24.1}
  g_{\mu\nu}\in C^{\hak+1,\sigma}(\mcM)
  \,,
\end{equation}
for some $\sigma\in[0,1)$, with  $\sigma \ge \lambda$ in the $\lambda$-H\"older case.

\input{firstStep}

\input{secondStep}

\input{thirdStep}

%% file: RemarkDiff.tex
\begin{remark}
\label{R30VI24.113} 
In our gluing results we allow only a finite number $\bluek$ of transverse derivatives. In the current section  $\bluek=\infty$ is allowed, because  equations  \eqref{15IX23.1a}, \eqref{5IV24.3} and \eqref{5V24.3}
can be  understood in the sense of Borel summation.
However, it is not clear whether $\bluek=\infty$ would make sense in
\eqref{8VI23.2}; this is at the origin of our restriction $\bluek<\infty$ in theorems such as Theorem~\ref{T6V24.1v2}.
\qedskip
\end{remark}

%% file: firstStep.tex
\subsubsection{First coordinate transformation.}
\label{ss17IV24.1}

For $\cu$ near $0$,
 on $\cH_2$
 we set 
\begin{align}
   u(\cu,\cx^A)  &=
   \psizero(\cx^A) +
    \psi_1(\cx^A )
   \cu
     + \psi_2(\cx^A )
   \frac{ \cu^2}2
        + \ldots
    +
      \red{\psi_{\bluek+2}}(\cx^A )
   \frac{ \red{\cu^{\bluek+2}}}{(\bluek+2)!}
    \,,
     \label{15IX23.1a}
\end{align}
with functions
\begin{equation}\label{11V24.1}
 \mbox{
  $\psi_i \in \XS{\kgamma+2-2i} $,   and where $\psi_1 >0$.
  }
\end{equation}
Equation~\eqref{15IX23.1a} should be understood in the sense of Borel-summation when $\bluek=\infty$.

 We find
\begin{align}\label{5IV24.5}
  u|_{\{\cu=0\}}
   =  \psizero
    \in \XS{\kgamma +2}
   \,,
   \quad
  \partial_\cu u |_{\{\cu=0\}}
    =   \psi_1
    \in \XS{\kgamma }
    \,,
    \quad
    \ldots
    \,,
    \quad
  \partial_{\cu}^{\bluek+2} u|_{\{\cu=0\}}
  =
  \psi_{\bluek+2}
    \in \XS{\kgamma -2\bluek-2}
  \,,
\end{align}
and
{{
\begin{align}\label{5IV24.5xc}
  \forall\  i >  \bluek+2
  \qquad
  \partial_{\cu}^ i u|_{\{\cu=0\}}
  = 0
  \,,
\end{align}
in particular it holds that
\begin{align}\label{5IV24.5xd}
  \forall\  0 \le 2i \le  \hak +2
  \qquad
  \partial_{\cu}^ i u |_{\{\cu=0\}} \in \XS{\kgamma +2-2 i}
  \,.
\end{align}
}}
(Should one wish to minimise losses of differentiability of the transformed metric away from $\{\cu = 0\}$,  in \eqref{15IX23.1a} one could apply
to the coefficients $\psi_i$ suitable extension maps
so that $u$ is smooth away from $\{\cu =0\}$, while maintaining \eqref{5IV24.5}.
But this is irrelevant for the considerations to follow.)

On $\cH_2$ we replace the coordinates $(u,x^A)|_{\cH_2}$ by a new set of coordinates $(\cu,\cx^A=x^A )$, where $\cu|_{\cH_2}$ is defined by \eqref{15IX23.1a}, and we define the coordinates $(\cu,\checkr,\cx^A)$ away from $\cH_2$
by setting $\checkr = r$
 and 
 {{setting $\check{x}^A$ and $\check{u}$ to be constant along the flow of the null geodesics}}%
 \footnote{Since $\hak\ge 2\bluek\ge 2$, there still exists a class of geodesics which are uniquely defined by their initial data, even though the metric might be poorly differentiable in different  coordinates used, as long as the coordinates are $C^1$-related to the well-behaved ones.}
 orthogonal to the level sets of $\cu$ within the hypersurface $ \cH_2$ of \eqref{30III24.1}.
(We remark that imposing $\partial_\chr u =0$, which would vastly simplify what follows, is not compatible with \eqref{21III22.1.1n}-\eqref{19VII23.21} unless $\partial_\cA u=0$.)

We emphasise that the above construction automatically preserves the null form of the metric; in particular \eqref{18VIII23.f1}-\eqref{18VIII23.f2} hold.

From \eqref{20III221b} we obtain
{{
\begin{eqnarray}
 \fourg
   &= &
   - \underbrace{
    \alpha U_{\mtdu}^2
     }_{=:\check \alpha}
      d\mtdu^2
    -  2
    \underbrace{
     (\alpha U_{\mtdu} U_{\check C} - \nu_A U_{\check u}\delta^A_{\check C})
    }_{=:\check \nu_\chC}
    d\mtdu \, d\mtdx^C
+
 2
 \underbrace{
  ( -\alpha U_{\mtdr}U_{\mtdu}
  + \nur    U_{\mtdu} + \nu_A U_{\check{u}} X^A_{\check{r}})
  }_{=: \check \nu _\chr}
   d\mtdu \, d\mtdr
 \nonumber
\\
& &
 +
 \underbrace{
 (
  g_{CD}
    -\alpha U_{\check C}  U_{\check D}
    + 2 \nu_A\delta^A_{\check D} U_{\check C})
    }_{=:g_{\mtdC\mtdD}}
d\mtdx^C d\mtdx^D
 \,.
  \label{17IV24}
\end{eqnarray}
}}


It holds that:

\begin{enumerate}

\item
Using
\begin{equation}\label{11IV24.11-b}
  \partial _\cu g_{\mu\nu} |_{\{\cu=0\}}
  =
  U  _\cu  \, \partial_u g_{\mu\nu} |_{\{\cu=0\}}  \in
       \XS{\kgamma}
       \ ,
       \ldots
       ,
       \
        \partial^i _\cu  g_{\mu\nu} |_{\{\cu=0\}}  \in
       \XS{\kgamma +2 -2i}
       \,,
\end{equation}
%
Equations  \eqref{5IV24.5xd}-\eqref{17IV24} show that
\begin{equation}\label{11IV24.11+}
 \forall\
 0\le 2i\le \hak
 \qquad
  \partial^i_\cu g_{\cu\cu} \restHtwo
  \,,
  \
  \partial^i_\cu g_{\cu\cA} \restHtwo
  \in   \XS{\kgamma -2i}  \,,
  \
  \partial^i_\cu g_{\cA\cB} \restHtwo
  \in
       \XS{\kgamma +1-2i}
  \,.
\end{equation}

\item
Recall the definitions of $\ysqr$ in \eqref{20VII23.w1} and  of 
$z$ in \eqref{19VI24.2}: 
\begin{align}
    \ysqr(\cu,\chr,\cx^A) = U_{\check C}{U_{\check D}} (\Lambda^{-1})^{\check C}{}_{A}  (\Lambda^{-1})^{\check D}{}_{ E} g^{EA}
    \,,
    \quad
     z
  :=\frac{
  \big(
   1+g^{AB} (\Lambda^{-1})^{ \chC}{}_B{}  U_{\chC}\nu_A
    \big)^2}{\alpha+g^{AB}\nu_A\nu_B}
  \,.
     \label{14IV24.21}
\end{align}
Since $\Lambda^A{}_{\cA}|_{\cH_2} $ is the identity,
using 
\eqref{3V24.p0}-\eqref{3V24.p1} 
we find
\ptcheck{25VI;  together}
\begin{equation}\label{14IV24.3b}
   \partial^i_\cu(\ysqr)\restHtwo
   \,,
   \
   {\redc
   \partial^i_\cu z\restHtwo
   \,,
   }
   \
   \partial^i_\cu U_{\chr}\restHtwo\,,
   \
   \partial^i_\cu
   X^A_{\chr}\restHtwo \in
       \XS{\kgamma +1-2i}
 \,.
\end{equation}
Taking a $\chr$-derivative  of \eqref{14IV24.21} gives
\begin{align}
    \partial_{\chr}(\ysqr)\restHtwo &= \partial_{\chr}\big( U_{\check C}{U_{\check D}} \delta^{\check C}{}_{A} \delta^{\check D}{}_{ E} g^{EA}\big)
    \nn
    \\
    & =
    2
    \underbrace{U_{\check C,\chr}}_{\in \XS{\kgamma } } U_{\check D} \delta^{\check C}{}_{A}  \delta^{\check D}{}_{ E} g^{EA}
    + U_{\check C} U_{\check D} \delta^{\check C}{}_{A}  \delta^{\check D}{}_{ E} \underbrace{\partial_{\chr}g^{EA}}_{\in \XS{\kgamma } }
    \in \XS{\kgamma }
    \,,
\end{align}
with a similar calculation for $z$, 
where we used  the notation $f_{,\chr} := \partial_{\chr} f$, and of course all terms on the right-hand side are  evaluated at $\zsecN$.
We also used
 $\partial_r g_{\mu\nu}\restHtwo \in \XS \hak$ and \eqref{14IV24.3b} to estimate the last term:
\begin{equation}\label{19VI24.3}
  \partial_\chr g^{EA} =  \partial_r g^{EA} +  \partial_\chr u  \partial_u  g^{EA}  +  \partial_\chr x^B  \partial_B  g^{EA}
    \in
     \XS{\kgamma }
  \,.
\end{equation}
By induction over $i$ and $j$,
%
\begin{equation}
\forall \
 0\le j + 2i  \le \hak +1
 \qquad
    \partial_{\cu}^i\partial^j_{\chr} z\restHtwo\,,\partial_{\cu}^i\partial^j_{\chr} \ysqr \restHtwo
     \in
       \XS{\kgamma +1-j-2i}
\,,
 \label{17IV24.1-}
\end{equation}
which immediately implies
\begin{equation}
\forall
 \
 0\le j + 2i  \le \hak +1
 \qquad
\partial^i_\chu \partial^j_\chr U_{\chr}
\,,
 \
\partial^i_\chu \partial^j_\chr X^A_{\chr} \in \XS{\kgamma +1-j-2i}  \,.
 \label{17IV24.1b}
\end{equation}
The last line can be rewritten as
\ptcheck{30VI24}
\begin{align}
\forall \
&
i \ge 1 \,,\
 0\le j + 2i \le \hak+1
 \qquad
 \partial^{i-1}_\chu \partial^{j+1}_\chr U_{\chu}
\,,
 \
\partial^{i-1}_\chu \partial^{j+1}_\chr X^A_{\chu}
 \in \XS{\kgamma +1-j-2i}
    \nonumber
\\
   & \Longleftrightarrow
   \quad
\forall \
 j\ge 1\,, \
 0\le j + 2i  \le \hak
 \qquad
 \partial^{i }_\chu \partial^{j }_\chr U_{\chu}
\,,
 \
\partial^{i }_\chu \partial^{j }_\chr X^A_{\chu} \in
 \XS{\kgamma -j-2i}
     \,.
 \label{17IV24.2}
     \end{align}
Together with \eqref{11IV24.11+},
Equations \eqref{17IV24.1b}-\eqref{17IV24.2} translate into
\ptcheck{2VII}
\begin{align}\label{17IV24.3}
 \forall \
 0\le j + 2i \le \hak
 \quad
   \partial^j_\chr  \partial^i_\cu g_{\cA\cB} \restHtwo
  &\in \XS{\kgamma +1-j-2i}
  \,, \
  \nonumber\\
  \partial^j_\chr  \partial^i_\cu g_{\cu\cu} \restHtwo
  \,,
   \ \partial^j_\chr \partial^i_\cu g_{\cu\cA} \restHtwo
   & \in \XS{\kgamma -j-2i}
  \,.
\end{align}
%

\item Using $ \check \nu _\chr \equiv g_{\cu \chr} =   -\alpha U_{\mtdr}U_{\mtdu}
  + \nur    U_{\mtdu}  + \nu_A U_{\check{u}} X^A_{\check{r}} $
 we also  obtain
\begin{equation}\label{11IV24.12xdc}
 \forall \
 0\le j + 2i \le \hak
 \qquad
\partial^j_\chr \partial^i_\cu g_{\cu\chr} \restHtwo  \in
    \XS{\kgamma -j-2i}
  \,.
\end{equation}
\end{enumerate}

%% file: secondStep.tex
\subsubsection{Second coordinate transformation.}
\label{ss17IV24.2}

We need next to make a change of coordinates $x^A$ on $\secNtwoNew$. Note that the first coordinate transformation required moving the null hypersurface $\{u=0\}$ in spacetime, while a change of $x^A$'s  does not. Therefore the current step can be viewed as a ``gauge transformation''  of sphere data, while the previous one has a substantially different character.

 In order to exploit the equations so far, and to avoid an explosion of notation, we rename the coordinates $\chx^\mu$ of Section~\ref{ss17IV24.1} to  $x^\mu$, and denote the new coordinates to be constructed here again by $\chx^\mu$. Thus \eqref{20III221b}
applies with the   metric functions  satisfing, instead of  \eqref{26V24.1}, 
for $
 0\le j + 2i  \le \hak $, 
\begin{equation}\label{17IV24.5}
  \partial^j_ r  \partial^i_ u g_{AB} |_\zsecN
   \in \XS{\hak +1-j-2i}
  \,,
  \quad
   \partial^j_ r  \partial^i_ u \alpha |_\zsecN\,,
   \
  \partial^j_ r\partial^i_\cu \nu_r |_\zsecN  \,,
 \
  \partial^j_ r  \partial^i_ u \nu_A |_\zsecN
   \in \XS{\kgamma -j-2i}
  \,.
\end{equation}

The construction will invoke   a map $\Phi:=\Phi_1$,  which is the  $t=1$ solution of the
 {
following flow on the surfaces of constant $u$ and $r$: 

\begin{equation}
 \frac{d\Phi_t}{dt}(\cu,\cx) = X\big(\cu,\Phi_t(\cu,\cx)\big)
  \,,
  \qquad
  \Phi_0(\cu,\cx)=\cx
  \,,
  \label{5IV24.2n}
 \end{equation}
or, in coordinate notation
\begin{equation}
 \frac{d\Phi^A_t}{dt}(\cu,\cx^B) = X^A\big(\cu,\Phi^C_t(\cu,\cx^B)\big)
  \,,
  \qquad
  \Phi^A_0(\cu,\cx^B)=\cx^A
  \,,
  \label{5IV24.2}
 \end{equation}
}
 with $X^A|_{\cH_2}$
 of the form
\begin{equation}\label{5IV24.3}
  X^A(\cu,\cx^B) |_{\cH_2}= \os{X}{0}{}^A(\cx^B)
  +  \os{X}{1}{}^A(\cx^B)\cu
   + \ldots
  +  \os{X}{\oldkg+1}{}^A(\cx^B)\frac{\cu^{\oldkg+1}}{(\oldkg+1)!}
  \,,
\end{equation}
with vector fields $ \os{X}{i} \in \XS{\hak +1-2i }(\secN)$;
Equation~\eqref{5IV24.3} should be understood in the sense of Borel-summation when $\bluek=\infty$.
Thus
\begin{align}\label{5IV24.5sdf}
  X^A|_\zsecN
   & =
    \os{X}{0}{}^A \in \XS{\hak +1}
   \,, \quad
  \partial_\cu X{}^A|_\zsecN =  \os{X}{1}{}^A \in \XS{\hak -1}
   \,,
   &\ldots\,,
   \nonumber
\\
  \partial_{\cu}^{\oldkg+1} X^A|_\zsecN
  & =  \os{X}{\oldkg+1}{}^A \in \XS{\hak-2\bluek-1}
  \,.
\end{align}
This implies
\begin{align}\label{5IV24.5b}
  \Phi^A|_\zsecN   \in \XS{\kgamma +1}
   \,, \quad
 \partial_\cu \Phi^A|_\zsecN   \in \XS{\kgamma-1}
   \,, \quad
   \ldots
   \,,
   \quad
  \partial_{\cu}^{\bluek+1}\Phi^A|_\zsecN
   \in \XS{\kgamma -2\bluek -1}
  \,.
\end{align}
We set
\begin{align}
 u = \cu\,,\quad
 r = \chr\,,
 \quad
    x^A (\cu,\cx^B)   =
     \Phi^A (\cu,\cx^B)
    \,.
    \label{6IV24.1}
\end{align}
In particular
\begin{equation}
 U_{\mtdr}\equiv 0 \equiv   X^A_{\mtdr}
 \,.
\end{equation}
Equation \eqref{20III221b} becomes
\begin{eqnarray}
 \fourg
   &= &
  (g_{AB} X^A_{\mtdu} X^B_{\mtdu}  -  \alpha  + 2 \nu_A X^A_{\mtdu})
  \, d\mtdu^2 \nonumber
+ 2 (g_{AB} {\Lambda}^A_{\pt{A}\check C} X^B_{\mtdu}
   +\nu_A {\Lambda}^A_{\pt{A}\check C}  )
   \,  d\mtdu \, d\mtdx^C
 \nonumber
\\
& &
+ 2  \nu_r
 \,  d\mtdu \, d\mtdr
+ \ g_{AB} {\Lambda}^A_{\pt{A}\check C} {\Lambda}^B_{\pt{B}\check D}
    \, d\mtdx^C d\mtdx^D
 \,,
  \label{15IV24.41}
\end{eqnarray}
leading directly to
(the reader is referred to~\cite[Lemma~A.2]{BartnikChrusciel1} or~\cite{Sickel,Bourdaud}
for composition of maps in Sobolev spaces)
\begin{equation}\label{19IV24.41}
    \partial^i_\cu g_{\cA\cB} |_\zsecN
   \,,
   \    \partial^i_\cu g_{\cu\chr} |_\zsecN
  \in \XS{\kgamma -2i}
  \,,
  \
  \mbox{and}
  \
   \partial^i_\cu g_{\cu\chu} |_\zsecN
  \,, \
     \partial^i_\cu g_{\cu\cA} |_\zsecN
    \in \XS{\kgamma -1-2i}
  \,.
\end{equation}
Since $u$ and $x^A$ are $\chr$-independent, the $\chr$-derivatives
follow a pattern identical to \eqref{17IV24.5}, namely, for $j \ge 1$,
\ptcheck{2VII
\\--\\wan: yes to all, except $gur$, which is $0\le j+2i \le \hak $, but presumably not needed so let's not worry about this?}
\begin{eqnarray}\label{19IV24.41er}
 \forall \
  1\le j+2i \le \hak \quad
  &
     \partial^j_\chr  \partial^i_\cu g_{\cA\cB} |_\zsecN
  \in \XS{\kgamma +1-j-2i}
  \,,
   \ \partial^j_\chr  \partial^i_\cu g_{\cu\chr} |_\zsecN
  \in \XS{\kgamma -j-2i}
  \,,
   \nonumber
\\
 &
  \partial^j_\chr \partial^i_\cu g_{\cu\cu} |_\zsecN
  \,, \
  \partial^j_\chr  \partial^i_\cu g_{\cu\cA} |_\zsecN
    \in \XS{\kgamma -j-2i}
  \,.
\end{eqnarray}

%% file: thirdStep.tex
\subsubsection{Third coordinate transformation.}
\label{ss17IV24.3}

The last step is to adjust the radial coordinate $r$; this is  a coordinate change on $\NIold$, thus a gauge and not a deformation. Again, in order to exploit the equations so far  and to avoid an explosion of notation, we rename the coordinates $\chx^\mu$ of Section~\ref{ss17IV24.2} to  $x^\mu$, and denote the  coordinates to be constructed here  by $\chx^\mu$. It follows from \eqref{19IV24.41}- \eqref{19IV24.41er} that \eqref{20III221b}
applies  with metric functions  satisfying
\begin{equation}\label{17IV24.5a}
 \partial^i_ u \nu_r |_\zsecN\,,
   \
   \partial^i_ u g_{AB} |_\zsecN
   \in \XS{\kgamma -2i}
  \,,
  \quad
  \partial^i_u\alpha  |_\zsecN  \,,
 \
    \partial^i_ u \nu_A |_\zsecN
   \in \XS{\kgamma -1-2i}
  \,,
\end{equation}
and,
for $1\le j+2i \le \hak $ and $j\ge 1$,
\begin{eqnarray}\label{19IV24.41erx}
     \partial^j_r  \partial^i_u g_{AB} |_\zsecN
  \in \XS{\kgamma +1-j-2i}
   \,,
   \ \partial^j_r  \partial^i_u \nu_r |_\zsecN
   \,,
  \partial^j_r \partial^i_u \alpha |_\zsecN
  \,,
  \partial^j_r  \partial^i_u \nu_A |_\zsecN
    \in \XS{\kgamma -j-2i}
  \,.
  \phantom{xxx}
\end{eqnarray}

We define a function $\rho>0$  by
{{
\begin{equation}\label{19IV24.53}
\rho^{(n-1)} := \frac{\sqrt{\det g_{AB}}} {\sqrt{\det \zgamma_{AB}}}
 \,.
\end{equation}
}}
The function $\fr$ is defined as the value of $\rho$ at $r=r_2$
\begin{equation}\label{19IV24.54}
  \fr  = \rho |_\zsecN
  \,.
\end{equation}
This defines $\fr$ as a smooth function, in the topologies listed, of the deformation-and-gauge data.

We set
\begin{equation}\label{12IX23.2g}
  u = \mtdu\,,
  \
   x^A = \mtdx^A
  \,,
  \
  \chr
   :=  \rho
   \,.
\end{equation}
It follows from \eqref{17IV24.5a}-\eqref{19IV24.53} that
\begin{equation}\label{17IV24.6}
 \forall  \  
0  \le  j+2i \le \hak
\,,
\qquad \partial^j_ \chr \partial^i_ \cu r |_\zsecN
 \in
    \left\{
      \begin{array}{ll}
       \XS{\kgamma -2i}, & \hbox{$j=0$;} \\
       \XS{\kgamma +1-j-2i}, & \hbox{$ j>0$.}
      \end{array}
    \right.
%
\end{equation}
Equation \eqref{20III221b} gives
\begin{eqnarray}
 \fourg
   &= &
  (-  \alpha  + 2 \nur R_{\mtdu})
  \, d\mtdu^2
+ 2 (  \nu_A
 +   \nur R_{\cA})
   \,  d\mtdu \, d\mtdx^A
+ 2  \nu_r R_{\mtdr}
 \,  d\mtdu \, d\mtdr
+ \ g_{AB}
    \, d\mtdx^A d\mtdx^B
 ,
 \qquad
  \label{15IV24.41a}
\end{eqnarray}
and from what has been said we obtain, with the first line for  $0\le 2i  \le \hak$ and the remaining ones for $j\ge 1$,
 \ptcheck{2VII}
\begin{align}\label{19IV24.41a}
    \partial^i_\cu
 &
g_{\cA\cB} |_\zsecN
   \,,
\partial^i_\cu g_{\cu\chr} |_\zsecN
   \
  \in \XS{\kgamma -2i}
  \,,
     \partial^i_\cu g_{\cu\cA} |_\zsecN
    \in
    \XS{\kgamma -1-2i}
    \,,
  \quad
   \partial^i_\cu g_{\cu\chu} |_\zsecN
   \in \XS{\kgamma -2-2i}
    \,,
\\
   \forall \,
   1\le j
 &
 +2i \le \hak\,,\
\quad
 \partial^j_\chr   \partial^i_\cu g_{\cA\cB} |_\zsecN
  \in \XS{\kgamma +1-j-2i}
   \,,   \ \partial^j_\chr  \partial^i_\cu g_{\cu\chr} |_\zsecN
   \,,\,
  \partial^j_\chr  \partial^i_\cu g_{\cu\cA} |_\zsecN
    \in
    \XS{\kgamma -j-2i}
  \,,
    \\
   \forall \,
   1\le j
    &
 +2i \le \hak-1\,,\
  \
\,
  \partial^j_\chr \partial^i_\cu g_{\cu\chu} |_\zsecN
   \in \XS{\kgamma -1-j-2i}
  \,.
\end{align}
%
\input{FlowProposition}

\subsection{$\EPsiStar \,\fourg_2)_{AB}$}
 \label{ss6V24/1}

We are ready now to construct the desired field $\EPsiStar \,\fourg_2)_{AB}$.
For the purpose of the definition
\eqref{8VI23.2} we take $\fourg$ in \eqref{19VII23.11} to be $\fourg_2$.  We rename the coordinates $( \chr, \cx^A)$ of the last section to $(r,x^A)$, and for $\hak<\infty$
we set
\begin{equation}\label{5V24.3}
  \EPsiStar \fourg_2)_{AB}
  = \sum_{j=0}^\hak
   E_j(\partial^j_r  g_{AB} |_\tsecNtwo)\frac{(r-\fr)^j}{j!
   }\in \XN{\hak}
   \,,
\end{equation}
where   $E_j(\partial^j_r  g_{AB} |_\zsecN)=E_j(\partial^j_r  g_{AB} |_\zsecN)(r,x^A)$ are tensor fields
such that the Taylor expansion in $r$ at $r=\fr$ of $\EPsiStar \fourg_2)_{AB}$ coincides with that of
\begin{equation}\label{5V24.4}
  \sum_{j=0}^\hak
    \partial^j_r  g_{AB} |_\tsecNtwo \frac{(r-\fr)^j}{j!
   }
   \,.
\end{equation}
Such extension maps in  H\"older spaces are constructed     in the proof of  \cite[Corollary~3.2]{AndChDiss}. In Sobolev spaces the existence of such maps can be justified as follows: By~\cite[Theorem 6.4.4]{BerghLofstrom} the spaces $W^{\ell,p}(\secN)$ with $p\ge 2$ embed  in the Besov spaces $B^\ell_{p,p}(\secN)$. The  extension map given in \cite[4.4, p.~193]{Stein} gives the desired extension $\EPsiStar \fourg_2)_{AB}$  in $W^{\kgamma+1/p,p}(\mcN)\subset W^{\kgamma,p}(\mcN)$. For $p\in (1,2)$ one notices that $W^{\ell,p}(\secN)\subset W^{s,p}(\secN)=B^s_{p,p}(\secN)$ for any $\ell-1< s<\ell$, and the desired extension is then in $W^{s+1/p,p}(\mcN)$ for all $s<\kgamma$, again a subset of $ W^{\kgamma,p}(\mcN)$.

When $\hak=\infty$ we define $\EPsiStar \fourg_2)_{AB}$ replacing  the  sum \eqref{5V24.4} by its Borel summation.
%

%% file: FlowProposition.tex
Summarising, we have proved that we can apply a deformation-and-gauge transformation to a spacetime metric $g$ with  the right differentiability of the final metric to apply the implicit function theorem in the next section (recall that $\bluek$ denotes the number of transverse derivatives to be glued, and note the different ranges of $\lambda$ and $p$ here, as compared to Theorem~\ref{T6V24.1v2}, because we do not have to solve any elliptic equations in this section):

\begin{theorem}
 \label{T6V24.1} Assume that  $\fourg $ is in $C^{\hak+1,\sigma}(\mcM)$ where $\sigma \in [0,1]$, with $\sigma\ge \lambda\in [0,1]$ in the $\lambda$-H\"older setting and $p\in[1,\infty]$ in the $L^p$-Sobolev setting in Definitión~\ref{D12V24.1}.
Let $I$ be an interval containing zero, and
  let $\bluek\,,\, \hak\in \N\cup\{\infty\}$   satisfy
 \begin{equation}\label{7V24.1a}
    \hak/2 - 1 \ge   \begin{cases}
         \bluek  \,, & \mbox{in the H\"older case,}
         \\
          \bluek+(n-1)/2p \,, & \mbox{in the $L^p$-type Sobolev case, $p>1$.}
     \end{cases}
 \end{equation}
Given  a set of deformation-and-gauge fields
\begin{align}\label{5IV24.5aagain2}
  &
   \psizero
    \in \XS{\hak  +2}
   \,,
   \quad
    \psi_1
    \in \XS{\hak  }
    \,,
    \quad
    \ldots
    \,,
    \quad
  \psi_{\bluek+2}
    \in \XS{\hak -2\bluek-2 }
  \,,
\\
 \label{28IV24.1again2}
  &
  \os{X}{0}{}^A   \in \XS{\hak  +1}
   \,, \quad
 \os{X}{1}{}^A  \in \XS{\hak  -1}
   \,, \quad
   \ldots
   \,,
   \quad
 \os{X}{\bluek+1}{}^A
   \in \XS{\hak  -2\bluek-1}
  \,,
\end{align}
there exist
\input{Point1}
where $\tsecNtwo = \{ u = 0,\, r=\fr|_{u=0}\}$.
 $\Psi$  is a composition of a map satisfying
\begin{align}\label{6IV24.23}
  u|_\zsecN
   =  \psizero
   \,,
   \quad
  \partial_\cu u |_\zsecN
    =   \psi_1
    \,,
    \quad
    \ldots
    \,,
    \quad
  \partial^{\bluek}_\cu u|_\zsecN
  =
  \psi_\bluek
  \,,
\end{align}
where $\zsecN = \{ u = \psizero, \, r=r_2\}$,
and of a  $t=1$ solution of the flow
\begin{equation}
 \frac{d\Phi^A_t}{dt}(\cu,\cx^B) = X^A\big(\cu,\Phi^C_t(\cu,\cx^B)\big)
  \,,
  \qquad
  \Phi^A_0(\cu,\cx^B)=\cx^A
 \,, 
  \label{5IV24.2a4}
 \end{equation}
%
where
\begin{equation}\label{5IV24.3a4}
  X^A(\cu,\cx^B) = \os{X}{0}{}^A(\cx^B)
  +  \os{X}{1}{}^A(\cx^B)\cu
   + \ldots
  +  \os{X}{\bluek+1}{}^A(\cx^B)\frac{\cu^{\bluek+1}}{(\bluek+1)!}
 \,,
\end{equation}
in the sense of Borel-summation when $\bluek=\infty$,
followed by a redefinition of $r$.
The map $\Psi$
and the functions $\partial_u^i\fr $
 depend  smoothly upon the deformation-and-gauge fields in the topologies listed.
\qed
\end{theorem}

%% file: Point1.tex
a diffeomorphism $\Psi$ and a function   $\fr:I\times \secN \to \R^+$ satisfying
$$\partial_u^i\fr(u,\cdot)\in \XS{\hak  -2\bluek}
$$
 which
  bring  $\fourg_2$ to a Bondi form with: for
  $
  0\le j+2i\le \hak $ in the first two lines and   $
  0\le j+2i\le \hak-1 $ in the last one,
\begin{align}\label{6V24.21}%
 \partial^i_ u g_{ A B} |_\tsecNtwo
   \,,
   \
 & \partial^i_ u g_{ ur} |_\tsecNtwo
  \in \XS{\kgamma -2i}
  \,,
     \partial^i_ u g_{ u A} |_\tsecNtwo
    \in
    \XS{\kgamma -1-2i}
    \,,
  \quad
   \partial^i_ u g_{ u u} |_\tsecNtwo
   \in \XS{\kgamma -2-2i}
    \,,
\\
 \forall\   j\ge 1 \,,
\quad
 &
 \partial^j_r   \partial^i_ u g_{ A B} |_\tsecNtwo
  \in \XS{\kgamma +1-j-2i}
   \,,
  \
   \ \partial^j_r  \partial^i_ u g_{ ur} |_\tsecNtwo
   \,,\,
  \partial^j_r  \partial^i_ u g_{ u A} |_\tsecNtwo
    \in
    \XS{\kgamma -j-2i}
  \,,
    \\
    &
  \partial^j_r \partial^i_ u g_{ u u} |_\tsecNtwo
   \in \XS{\kgamma -1-j-2i}
  \,,
  \label{6V24.23}
\end{align}

%% file: NonlinearWanCharge.tex
\section{Radial charges}
 \label{s21V24.1}

In this section, we define $\kQ{1}{}$ and $\kQ{2}{}$, the linearisations of which constitute  gauge-invariant obstructions to the linearised gluing problem in the case $m\neq 0$.
Indeed, for metrics which asymptote to a Birmingham-Kottler metric as $r$ tends to infinity, the right-hand sides of the $r$-derivatives of  $\kQ{1}{}$ and $\kQ{2}{}$  are at least quadratic in the deviation between $\fourg$ and its asymptotic Birmingham-Kottler counterpart. Therefore their linearisations are radially conserved. These  linearisations  of $\kQ{1}{}$ and $\kQ{2}{}$ coincide with their counterparts in~\cite{ChCong1,ChCongGray1}, and are therefore invariant under linearised gauge transformations. This implies that deformations and gauge transformation of $\kQ{1}{}$ and $\kQ{2}{}$ which are of order $\epsilon$ lead to transformations of $\kQ{1}{}$ and $\kQ{2}{}$ which are of order $\epsilon^2$.

\subsection{$\protect\kQ{1}{}$}

Using the Einstein  equations \eqref{22XII22.2} and \eqref{22XII22.3}, it can be verified that the following transport equation holds (cf.\ Appendix \ref{app6V24.1}):
\begin{align}
    \partial_r \left[r^{n+1} e^{-2\beta}\zhTBW_{AB}(\partial_r U^B)+ 2 r^{n-1}\spaceD_A\beta\right]
      &=
      \frac{r^n}{2(n-1)}\spaceD_A\bigg[\zhTBW^{EC}\zhTBW^{BD} (\partial_{r} \zhTBW_{EB})(\partial_r \zhTBW_{CD})\bigg]
      \nn
      \\
      &\quad
      -r^{n-1}\zhTBW^{EF} \spaceD_E (\partial_r \zhTBW_{AF})\,.
      \label{23II24.21}
\end{align}
Denoting the term in the square brackets of the left-hand side as
\begin{equation}
    \overadd{*}{H}_{uA}:=r^{n+1} e^{-2\beta}\zhTBW_{AB}(\partial_r U^B) + 2 r^{n-1}\spaceD_A\beta\,,
\end{equation}
the obstructions $\kQ{1}{}(\pi^A)$ are defined as a family of maps, parameterised by the $\pi^A$'s,  
 on the space of Bondi cross-sectional data, given by the projection of the above onto the space of Killing vectors of $\ringh$: for vector fields $\pi_A$ satisfying $\zspaceD_{(A}\pi_{B)} = 0$ on $\secN$
and $x\in\CSdataBo{\secN, \bluek;\hak}$,
\begin{align}
    \kQ{1}{}(\pi^A)[x]:= \int_{\secN} \pi^A \overadd{*}{H}_{uA}
    \,\fullmeasure
    \,,
\end{align}
where $\fullmeasure=\sqrt{\det \gamma_{AB}}\,d^{n-1}x $ is the natural measure on  $\secN$ induced by the spacetime metric $\fourg$.

\subsection{$\protect\kQ{2}{}$}
Again from the Einstein equations (cf.\ Appendix \ref{app6V24.1}), one can verify that we have the following transport equation:
\begin{align}
    &(n-1)\partial_r
    \bigg(
    \underbrace{r^{n-3} V - \frac{r^{n-2}}{n-1}D^A\partial_r(r^2U_A) - \frac{2 r^{n-2}}{n-1}  e^{2\beta} \Delta \beta}_{=:  \chi}
    \bigg)
    \nn
    \\
    &=
    2 r^{n-1} [D_B,\partial_r] U^B + \partial_r( 2 r^{n-2}  D^A(e^{2\beta}) D_A \beta)
    \nn
    \\
    &\quad + 2 r^{n-2}   [D^A,\partial_r](e^{2\beta} D_A \beta) - 2 r^{n-1}
 D^A[\partial_r(e^{2\beta}/r)D_A\beta]
    \nn
    \\
    &\quad
    +r^n[D_B,\partial_r]\partial_r U^B
    - 2\zhTBW^{AB}e^{2\beta} r^{n-3}\spaceD_A \spaceD_B \beta
    \nn
    \\
    &\quad
    +
    e^{2\beta}r^{n-3}\bigg[-2 \Lambda  r^2 +
                 R[\zhTBW]
                -2\zhTBW^{AB}  (\spaceD_A\beta) (\spaceD_B \beta)
                 -\frac{1}{2}r^4 e^{-4\beta}\zhTBW_{AB}(\partial_r U^A)(\partial_r U^B)\bigg]
                 \nn
    \\
    & \quad
    - 2 r^{n}D_B [(\partial_r U^B) \partial_r\beta]
        + r^{n}D^A \left[(\partial_r\zhTBW_{AB})(\partial_r U^B)\right]
        \nn
       \\
       &\quad
       -  D^A\left[\frac{e^{2\beta}r^{n-1}}{2(n-1)}D_A\bigg(\zhTBW^{EC}\zhTBW^{BD} (\partial_{r} \zhTBW_{EB})(\partial_r \zhTBW_{CD})\bigg)\right]
      +r^{n-2} D^A[ e^{2\beta}\zhTBW^{EF} \spaceD_E (\partial_r \zhTBW_{AF})]
                                 \,.
                                 \label{6V24.31}
\end{align}
Given $x\in \CSdataBo{\secN, \bluek;\hak}$ the obstruction $\kQ{2}{}$ is defined as: 
\begin{align}
    \kQ{2}{}
    [x] := \int_{\secN} e^{-2\beta}\chi
    \fullmeasure
    \,.
\end{align}

\subsection{Further radial charges}

When the mass parameter $m $ vanishes, further radial charges with similar properties have been listed in~\cite{ChCong1,ChCongGray1}. Nonlinear counterparts of these linearised radial charges can be obtained by, e.g., replacing in the definitions of~\cite{ChCong1,ChCongGray1} the linearised metric perturbations $\delta g_{\mu\nu}$ by $g_{\mu\nu} -\zfourg_{\mu\nu} $. We will collectively denote this set of radial charges at $r_a$ as $Q[\cdot]$. The key properties of these radial charges, as relevant for our problem at hand, are:

\begin{enumerate}

  \item Suppose that a set of null hypersurface data $y\in \CdataBo{\mcN, \bluek;\hak}$ satisfies $y-\yzero = O(\epsilon)$, then~\cite{ChCong1,ChCongGray1} 
  \begin{equation}\label{17V24.1}
    \partial_r 
    \big( Q[y|_{r }]
     \big)= O(\epsilon^2)
    \,,
  \end{equation}
  where the explicit formulae for $\partial_r  \kQ{1}{}$ and $\partial_r\kQ{2}{}$ can be obtained from the equations in the previous sections.
  We can define  a \textit{charge-transport map} $TQ$ by integration:
  \begin{equation}\label{17V24.2}
   TQ[y] : =Q[y|_{r=r_1}] + \int_{r_1}^{\fr} \partial_r Q[y]  dr
   =Q[y|_{r=r_1}] +O(\epsilon^2)
    \,.
  \end{equation}

  \item Given a metric $\fourg$, let us denote by $z^*(x,\fourg)$ the action of   a set of gauge-and-deformations  $z\in \GandD{\secN, \bluek;\hak}$ (cf.\ Definition \ref{D12V24.1} and {{Theorem \ref{T6V24.1}}}) on a
   codimension-two data set $x\in \CSdataBo{\secN, \bluek;\hak}$. If $z= O(\epsilon)$, then~\cite{ChCong1,ChCongGray1}
  \begin{equation}\label{17V24.3}
     Q[z^*(x,\fourg)]-Q[x]= O(\epsilon^2)
    \,.
  \end{equation}
\end{enumerate}

%% file: argument2a.tex
\section{The gluing up to radially-conserved charges}
\label{s10V24.1}

As a step  to prove Theorem~\ref{T24VIII23.1}, we establish  a
nonlinear-gluing result up-to-radial obstructions.
Indeed, it turns out  that the gluing-problem of order $\bluek$ near  Birmingham-Kottler metrics  can always  be solved up to a finite-dimensional space of obstructions determined by $\CSdataBo{\secN_1, \bluek;\hak}$.

%% file: NewTheoremv3.tex
\begin{lemma}[Gluing up to radial obstructions]
 \label{L6V24.1v3}
\input{Regularity3v2a}%
Let $r_1<r_2$,
and for $a=1,2$, let $\mathring x_a$ be codimension-two data  arising from a Birmingham-Kottler metric $\zfourg$ at $\{u=0,r_a\}$.
There exist

\begin{enumerate}
\item
  a finite set of radial charges $Q$, and

    \item a neighborhood $\mcU$ of $\zfourg$ in the space of $C^{\hak+1,\sigma}$ metrics defined near $r=r_2$, and

\item    neighborhoods $\mcO_a\subset \CSdataBo{\secN_a, \bluek;\hak}$   of $\mathring x_a$, and

\item a smooth  map $\OneMoreMap_\Phi$  from $\mcO_1\times \mcO_2\times \mcU$ to the set of characteristic data $\CdataBo{\mcN,\bluek;\hak}  $, and

\item a smooth  map $\OneMoreMap_\mcG$  from $\mcO_1\times \mcO_2\times \mcU$ to the deformation-and-gauge data $ \GandD{\mcN, \bluek;\hak} $,

\end{enumerate}
such that the following holds: Given two   codimension-two data sets   $x_{r_a}\in \mcO_a$, with $x_{r_2}$ induced by a metric $\fourg_2\in \mcU$,   the  vacuum characteristic data set $\OneMoreMap_\Phi(x_{r_1}, x_{r_2},\fourg_2)$
 
\begin{enumerate}
  \item[a)]
is compatible with $x_{r_1}$ and
\item[b)]
induces   a deformed codimension-two data set
 $z_{r_2}^*(x_{r_2},\fourg_2)$,
  where $z_{r_2} =  \OneMoreMap_\mcG(x_{r_1},x_{r_2},\fourg_2)$, if and only if
\begin{equation}\label{18V24.1}
  Q[z_{r_2}^*(x_{r_2},\fourg_2)]
    =
   TQ\big[
    \OneMoreMap_\Phi(x_{r_1},x_{r_2},\fourg_2)
  \big]
  \,,
\end{equation}
where $TQ$ is as in \eqref{17V24.1}.
\end{enumerate} 
\end{lemma}
 
In other words, we can use the map $\OneMoreMap_\Phi$ to solve the gluing problem if we can arrange that the finite number of conditions \eqref{18V24.1} is satisfied. We will show how to do this in the situation considered in the next section. 

%% file: Regularity3v2a.tex
Let $\bluek \in \N$, $\hak\in \N\cup\{\infty\}$, 
$\sigma\in[0,1)$, $\sigma \ge \lambda \in (0,1)$ in the $\lambda$-H\"older case, $p\in (1,\infty)$ in the $L^p$-type Sobolev case. 
   We suppose that, in $n$-space dimensions, $n \ge 3$, the regularity index  $\hak$ satisfies
\begin{equation}\label{n25V24.11Xabc}
 \hak
  \left\{
    \begin{array}{ll}
     \ge  2+2\bluek
    &  \mbox{in the H\"older case, or}
 \\ > 2+2\bluek +(n-1)/p \
   &  \mbox{in the $L^p$-type Sobolev case.}
    \end{array}
  \right.
\end{equation}

%% file: NonlinearProofWan.tex
\begin{remark}
\label{R12V24.1}
When the mass parameter   of $\zfourg$ vanishes, the number of radial charges  is given in the last line of Table~\ref{T26XI22.1}
  in spacetime dimension four and $k=2$; 
   see~\cite[Tables~1.1-1.3]{ChCongGray1} for the general case.
 When the mass parameter of $\zfourg$ is non zero, the number of radial charges equals $c_\zgamma+1$, where $c_\zgamma$ is the dimension of the space of Killing vectors of  $(\secN,\zgamma)$, with the radial charges $Q=(\kQ{1}{},\kQ{2}{})$ given by the integrals of Section~\ref{s21V24.1}.
\input{ChargeTable}
 \qedskip
\end{remark}

\proof
\input{Notation}

Next, we define $\OneMoreMap (x_{r_1},x_{r_2},\vphi{i},z_{r_2},\fourg_2)$ as the characteristic data set compatible with $x_{r_1}$,  with the characteristic data field $\gamma_{AB}$ 
given by
\begin{equation}\label{8VI23.2x}
   r^2\gamma_{AB}
   =\notchi
     \Big(
     \underbrace{
    \zfourg _{AB}
     +
    \phi_1\big((\Psionestar \fourg_1)_{AB} -
    \zfourg _{AB}
     \big)
    +
     \phi_2
     \big(\EPsiStar    \fourg_2)_{AB} -
    \zfourg _{AB}
    \big)
    +
    \sum_{  i\in\iota_{\ralpha,m}  }\kappa_i \, \vphi{i}_{AB}
     }_{=:\hat{g}_{AB}}
     \Big)
     \,,
\end{equation}
where $\omega$, $\phi_1$, $\phi_2$, $\iota_{\ralpha,m}$, $\kappa_i$,  and $\vphi{i}$ have been defined below \eqref{8VI23.2}, and

\begin{enumerate}
  \item  $\fourg_a$, $a=1,2$, are $C^{\hak+1}$
   vacuum metrics inducing $x_{r_a}$,
  and
  \item $\EPsiStar \,\fourg_2)_{AB}$ is constructed from $z_{r_2}\in \GandD{\mcN, \bluek;\hak} $ and from $\fourg_2$ using Theorem~\ref{T6V24.1} and \eqref{5V24.3}-\eqref{5V24.4}.
\end{enumerate}
Given 
two codimension-two data sets $x_{r_1}$ and $x_{r_2}$ of order $\bluek$ and the metric $\fourg_2$,  we wish to find $(\vphi{i},z_{r_2})$ solving the equation
\begin{equation}\label{24V24.1}
    \Mainmap_{x_{r_1}}\big(
   \OneMoreMap (x_{r_1},x_{r_2},\vphi{i},z_{r_2},\fourg_2)
   \big) =
  z_{r_2}^*(   x_{r_2},\fourg_2)
  \,,
\end{equation}
using the implicit function theorem.

\input{linear}

It holds that:

\begin{enumerate}
  \item For $r\in [r_1+2\mepsilon, \fr-2\mepsilon ]$  we have
\begin{equation}\label{8VI23.4}
  g_{AB}
   =\notchi
     \Big(
    \zfourg _{AB}
    +
     \sum_{  i\in\iota_{\ralpha,m}  }\kappa_i \, \vphi{i}_{AB}
     \Big)
     \,.
\end{equation}
Tracelessness of the $\vphi{i}_{AB}$'s shows that
the determinant  $\det \hat g_{AB}  $ is a polynomial in the  $\vphi{i}_{AB}$'s without linear terms. It follows that the linearisation of   $\notchi$, as given by \eqref{22VIII23.1}, with respect to the  $\vphi{i}_{AB}$'s is zero.

  \item The linearisation of the map defined by the second coordinate transformation of Section~\ref{ss17IV24.2} corresponds to the linearised gauge-transformations $\partial_u^i \xi^A$ of \cite{ChCong1,ChCongGray1}. For example, let $\Psi$ be generated by a vector field $\zeta^A\partial_A$. Using the formula
$$
 \det \big( g_{AB} + \epsilon A_{AB}\big)
  =  (\det  g_{AB})  \big(1 + \epsilon g^{AB}A_{AB} +  O(\epsilon^2)\big)
 $$
 one finds, on $\secNtwo$,  that the linearisation of $(\Psi^*g)_{AB}$ with respect to $\Psi$ at $
 \Psi=\text{Id}$  is
 $$
  C(\zeta)_{AB}:= D_A \zeta_B + D_B \zeta_A - \frac{2g^{CD} D_C \zeta_D}{d}
     g_{AB}
    \,,
  $$
  where $d=n-1$. These are the linearised gauge transformations $\zeta^A \partial_A$ of \cite{ChCong1,ChCongGray1}.

\item Similar calculations show that the linearisation of the map defined by the first coordinate transformation of Section~\ref{ss17IV24.1} corresponds to the linearised gauge-transformations $\partial_u^i \xi^u$ of \cite{ChCong1,ChCongGray1}.

\end{enumerate}

Consider the image, say $\obstr$,   of the linearisation with respect to its first two arguments of the map
\begin{equation}\label{25V24.1}
 (\vphi{i},z_{r_2},x_{r_1},x_{r_2},\fourg_2) \mapsto \Mainmap_{ x_{r_1}}
  \big(
 \OneMoreMap (x_{r_1},x_{r_2},\vphi{i},z_{r_2},\fourg_2)
 \big) -
  z_{r_2}^*(  \red{ x_{r_2}},\fourg_2 )
  \,.
\end{equation}
 at $(0,0,\mathring{x}_{r_1},\mathring{x}_{r_2},\zfourg)$.
By%
\footnote{In \cite{ChCong1} and \cite{ChCongGray1} $L^2$-based Sobolev spaces are considered, with stronger $r$-differentiability hypotheses than here in~\cite{ChCong1}. But the analysis in both references applies without further due to the $\XS{}$ and $\XN{}$ spaces used here.}
\cite[Theorem~5.1]{ChCong1} in spacetime dimension four, or by \cite[Theorem~6.1]{ChCongGray1} in higher dimensions,

\begin{enumerate}
  \item this  linearisation  is surjective on $\obstr$,
  \item with splitting kernel,
  say  $\mathfrak{K}$,
  and
  \item letting $q$ be the dimension of the space of radial charges (cf.~Remark~\ref{R12V24.1}), near $\mathring x_2$ we can write
      \begin{equation}\label{25V24.2}
           \CSdataBo{\secNtwo, \bluek;\hak}
             = \obstr\oplus \R^q
           \,.
      \end{equation}
\end{enumerate}

Returning to the proof of Lemma~\ref{L6V24.1v3}, let $\Pi$ denote the projection on the first factor in \eqref{25V24.1}.
The  implicit function theorem (cf., e.g., \cite[Theorem~5.9]{LangFundamentals}) shows that there exist  neighborhoods $\mcU$ and $\mcO_a$ as in the statement of Lemma~\ref{L6V24.1v3} and
unique fields $(\vphi{i},z_{r_2})$,
belonging to the closed subspace complementing the kernel $\mathfrak{K}$, such
that the equation
%
\begin{equation}\label{24V24.3}
 \Pi\Big( \Mainmap_{x_{r_1}}
 \big(
 \OneMoreMap (x_{r_1},x_{r_2},\vphi{i},z_{r_2},\fourg_2)
 \big)
 - z_{r_2}^*(   x_{r_2},\fourg_2)
  \Big) =0
\end{equation}
holds.
The maps $\OneMoreMap_\Phi$ and $\OneMoreMap_\mcG$  are  defined as
\begin{equation*}
   \OneMoreMap_\Phi(x_{r_1},x_{r_2},\fourg_2) :=
 \OneMoreMap (x_{r_1},x_{r_2},\vphi{i},z_{r_2},\fourg_2)
   \,,
   \quad
   \OneMoreMap_\mcG(x_{r_1},x_{r_2},\fourg_2):= z_{r_2}
   \,,
\end{equation*}
where  $(\vphi{i},z_{r_2})$ are the fields just mentioned. The proof for finite $\hak$ is completed.

Finally, for each $k\in\N$ the fields $(\vphi{i},z_{r_2})$ are obtained by 
{
using the implicit function theorem based on the elliptic system of equations of~\cite{ChCongGray1},
with a unique solution,  
which implies in particular that the solution is independent of $\hak \in \N$ satisfying \eqref{n25V24.11Xabc}. A standard argument  justifies then 
}
the claim for $\hak=\infty$.
\qedskip

%% file: ChargeTable.tex
\begin{table}[t]
  \centering
  \begin{tabular}{||c|c|c|c||}
  \hline
  \hline
      & $S^2$ & $\T^2$ &   genus $\genus\ge 2$\\
  \hline
    $\kQ{1}{}$: $m=0$
        & 6
                &  2
                     & 0
\\
    \phantom{$\kQ{1}{}$,} $m\ne 0$
        & 3
                &  2
                     & 0
\\
  \hline
    $\kQ{2}{}$:  $m=0$
        & 4
                &  1
                     & 1
\\
    \phantom{$\kQ{2}{}$:}  $m\neq0$
        & 1
                &  1
                    & 1
\\
  \hline
    $\kQ{3,1}{}^{[\harm]}$: $m=0$
        & 0
                & coincides with  $\kQ{2}{}$
                     & $2\genus $
\\
     \phantom{$\kQ{3,1}{}^{[\harm]}$: } $m\neq0$
        & 0
                &  0
                     & 0
\\
  \hline
    $q_{AB}^{[\TTt]}$ : $m=0$, $\ralpha =0$
        & 0
                &  2
                     & $6(\genus -1)$
\\
  \phantom{$q_{AB}^{[\TTt]}$ :} $m=0$, $\ralpha \ne 0$
        & 0
                &  0
                     & 0
\\
     \phantom{$q_{AB}^{[\TTt]}$ : } $m\neq0$
        & 0
                &  0
                     & 0
\\
  \hline
    $\kQ{3,2}{}^{[\harm]}$: $m=0$
        & 0
                & 0
                     & $2\genus $
\\
     \phantom{$\kQ{3,1}{}^{[\harm]}$: } $m\neq0$
        & 0
                &  0
                     & 0
\\
  \hline
 $\overset{[2]}{q}{}_{AB}^{[\TTt]}$: $m=0$
        & 0
                &  2
                     & $6(\genus -1)$
\\
     \phantom{$\overset{[2]}{q}_{AB}^{[\TTt]}$:} $m\neq0$
        & 0
                &  0
                     & 0
  \\
  \hline
     together: $m=0$, $\ralpha =0$
        & 10
                &   7
                     & $16\genus  -11$
\\
   \phantom{together:}  $m=0$, $\ralpha \ne 0$
        & 10
                &  5
                     & $10\genus - 5$
\\
    \phantom{together:} $m\ne 0$
        & 4
                &  3
                     & 1
\\
  \hline
  \hline
\end{tabular}
  \caption{The dimension of the space of obstructions   for linearised $\Ctwo$-gluing, spacetime dimension four, from~\cite{ChCong1}.
On $S^2$ the four obstructions associated with $\kQ{2}{}$ correspond to spacetime translations, the three obstructions associated with $\kQ1{}$ when $m\ne0$ correspond to rotations of $S^2$, with the further three obstructions arising when $m=0$ corresponding to boosts. The reader is referred to~\cite{ChCong1} for further definitions.}
  \label{T26XI22.1}
\end{table} 

%% file: Notation.tex
In order to proceed some  notation will be useful. Given an element $x_{r_1}\in   \CSdataBo{\secNone, \bluek;\hak} $
and a  function  $\fr>0 $ on $I\times \secNtwo$
 let us denote by $\MainMap_{x_{r_1}}$
 {{the map which, to a set of characteristic data
 $y\in  \CdataBo{\mcN, \bluek;\hak} $ compatible with $x_{r_1}$, assigns a codimension-two data set 
 $ \MainMap_{x_{r_1}}(y)  \in   \CSdataBo{\tsecNtwo, \bluek;\hak} $; specifically, the data set $ \MainMap_{x_{r_1}}(y) $ is obtained by the restriction of the fields on $\mathcal{N}$, produced by acting the map $\Xi$ of Theorem \ref{T24VI23.1} on $y$, onto $\tsecNtwo$.}}

%% file: linear.tex
\begin{remark}
 \label{R30VI24.1}
A comment concerning the integration range in $r$ might be in order, as here the gluing takes place at $r=\fr(u=0,x^A)$.
 The question then arises, whether this affects the relevance to the current work of the linearised equations analysed in~\cite{ChCong1,ChCongGray1}, where the gluing takes place at $r=r_2$. We assert that the results in these last two references apply without further due.

To see this, consider a  family of spacetime metrics parameterised by a parameter $\epsilon$. Let $F(\epsilon, r, u,x^A)$ denote a collection of fields, built from the metric functions
 and their derivatives,
   which satisfies a transport equation of the form
\begin{equation}\label{27VI24.1}
  \partial_r F(\epsilon,r,\cdot) = f (\epsilon, r,\cdot)
  \,,
\end{equation}
and  such that $F|_{\epsilon=0}$ takes the Birmingham-Kottler values. Let $\fr(\epsilon,\cdot)$ be a family of functions such that $\fr=r_2$ at $\epsilon=0$.
The gluing equations here take the form
\begin{equation}\label{27VI24.2}
 F(\epsilon,\fr(\epsilon,\cdot),\cdot) =  F(\epsilon,r_1,\cdot) +\int_{r_1}^{\fr(\epsilon,\cdot)} f (\epsilon,s)\, ds
  \,.
\end{equation}
Differentiating with respect to $\epsilon$ and setting, as usual, $\delta F = \frac{\partial F}{\partial \epsilon}\big|_{\epsilon=0}$  one obtains
\begin{equation}\label{27VI24.2xd}
 \delta F(r_2,\cdot)
  +
   \Big(
   \frac{\partial F(\epsilon,r_2,\cdot) }{\partial r}
   \frac{\partial \fr(\epsilon,r_2,\cdot) }{\partial \epsilon}
  \Big)\Big|_{\epsilon=0}
   =  \delta F(r_1,\cdot) +\int_{r_1}^{r_2} f (\epsilon,s)\, ds
 + f (0,r_2 )
   \frac{\partial \fr(\epsilon,r_2,\cdot) }{\partial \epsilon}\Big|_{\epsilon=0}
  \,.
\end{equation}
which is equivalent to
\begin{equation}\label{27VI24.2xde}
 \delta F(r_2,\cdot)
   =  \delta F(r_1,\cdot) +\int_{r_1}^{r_2} f (\epsilon,s)\, ds
  \,,
\end{equation}
which are the equations analysed in the $\fr=r_2$--linearisation procedure in~\cite{ChCong1,ChCongGray1}.
\qedskip
\end{remark}

%% file: argument2.tex
\section{The gluing to a nearby Kottler-(A)dS  metric}
\label{s25V24.11}

We are ready now to pass to our main result:

\input{MainTheorem}

\proof
The result follows in a standard way from Lemma~\ref{L6V24.1v3}, see~\cite{ACR2}, or \cite{CorvinoSchoen2,ChDelay} in a related context.
The only thing to check is that the families listed contain the whole set of compensating charges. In spacetime-dimension four this has been shown in \cite[Section~6]{ChCong1}.
For the Myers-Perry metrics near the Birmingham-Kottler metrics this has been shown in the linearised case in \cite[Section~7]{ChCongGray1},
which suffices for the purpose of our small-deformation results here.
For negatively curved (i.e., $R(\zgamma) <0$) Birmingham-Kottler metrics the radial charge $\kQ{1}{}$ is trivial, as the relevant metrics $\gamma_{AB}$ have no Killing vectors, and so only the mass parameter remains.

 For illustration we give the proof which covers the last case,  when the dimension of the space of charges is one, i.e.\ the only obstruction is the mass parameter, as then the argument is completely elementary, and proceeds as follows: Suppose  that $x_{r_1}$ is $\epsilon$-near to, e.g., a negatively curved Birmingham-Kottler metric $\fourg[\mathring m]$  with mass parameter $\mathring m$. We can normalise $ \kQ{2}{}$ so that for a  codimension-two data set, say $\mathring x_{m,r}$, induced by a Birmingham-Kottler metric $\zfourg[m]$ on the level sets of $r$ within $\{u=0\}$, we have
\begin{equation}\label{25V24.31a}
  \kQ{2}{}
  [\mathring x_{m,r}] = m
  \,.
\end{equation}
For any $z_{r_2}$ which is $\epsilon$-small we have, in view of \eqref{17V24.3},
\begin{equation}\label{25V24.31}
  \kQ{2}{}
  [z_{r_2}^*(\mathring x_{m,r_2})]= m + O(\epsilon^2)
  \,.
\end{equation}
There exists a constant $C>0$ such that
\begin{equation}\label{25V24.32}
  |\kQ{2}{}
  [ x_{r_1}] - \mathring m |\le C \epsilon
  \,.
\end{equation}
Given $s\in [-2C \epsilon, 2C \epsilon]$, Lemma~\ref{L6V24.1v3} provides characteristic data
$$
 y_{\mathring m+s}:=\OneMoreMap_\Phi(x_{r_1},\mathring x_{r_2,\mathring m+s},\fourg[\mathring m+s])
$$
connecting $x_{r_1}$ and
 $z_{r_2}^*(\mathring x_{\mathring m +s,r_2},\fourg[\mathring m+s])$ such that (cf.~\eqref{17V24.3})
  \begin{equation}\label{25V24.33}
   T\kQ{2}{}
   [ y_{\mathring m+s}]
   =\kQ{2}{}
   [x_{ r_1}] +O(\epsilon^2)
    \,.
  \end{equation}

Consider, now, the continuous function
\begin{equation}\label{18V24.1a}
 [-2C \epsilon, 2C \epsilon]\ni  s\mapsto F(s):=
 \underbrace{ \kQ{2}{}
   \big[
   z_{r_2}^*(\mathring x_{\mathring m +s,r_2},\fourg[\mathring m+s])
   \big]
   }_{=\mathring m + s + O(\epsilon^2)}
   -
   \underbrace{
   T\kQ{2}{}
    \big(
    y_{\mathring m+s}
  \big)
  }_{\in \mathring m +[-C\epsilon,C\epsilon]}
  \,.
\end{equation}
We have
$$
F(-2C\epsilon)\le  -C\epsilon + O(\epsilon^2)
 \ \mbox{and} \
F(2C\epsilon)\ge   C\epsilon + O(\epsilon^2)
\,.
$$
Continuity implies that, for $\epsilon$ small enough,  there exists $s$ such that $F(s)=0$, which provides the desired codimension-two data set induced by the metric $\fourg[\mathring m+s]$.
\qed

%% file: MainTheorem.tex
\begin{theorem}[Gluing to a nearby metric]
 \label{T6V24.1v2}
Let $r_1,r_2\in \R$ with $0<r_1<r_2$.
\input{Regularity3}%
Let $x_{r_1}\in \CSdataBo{\secN_1, \bluek;\hak}$ be  a  codimension-two Bondi data  set   sufficiently near to the data arising from one of the following $(n+1)$-dimensional metrics  with nonzero mass:
\begin{equation*}\label{4V24.1v2}
   \left\{
     \begin{array}{ll}
       \mbox{Kerr-(A)dS metrics}, & \hbox{when $\Lambda \in \R$, $\secN\approx S^{n-1}$ or a quotient thereof;} \\
       \mbox{Birmingham-Kottler metrics}, & \hbox{when $\Lambda\in \R$, $R(\zgamma) <0$.}
     \end{array}
   \right.
\end{equation*}
There exist
a function $\fr>0$ and
a  null-hypersurface data set $y\in\CdataBo{\mcNnew , \bluek;\hak}$ connecting $x_{r_1}$ with a codimension-two data set at $\{r=\fr\}$ induced by a nearby metric within the corresponding family.
\end{theorem}

%% file: Regularity3.tex
Let $\bluek \in \N$, $\hak\in \N\cup\{\infty\}$, with 
  $ \lambda \in (0,1)$ in the $\lambda$-H\"older case, or $p\in (1,\infty)$ in the $L^p$-type Sobolev case,  in the Definition~\ref{D12V24.1} of the function spaces.
   We suppose that, in $n$-space dimensions, $n \ge 3$, the regularity index  $\hak$ satisfies
\begin{equation}\label{n25V24.11X}
 \hak
  \left\{
    \begin{array}{ll}
     \ge  2+2\bluek
    &  \mbox{in the H\"older case, or}
 \\ > 2+2\bluek +(n-1)/p \
   &  \mbox{in the $L^p$-type Sobolev case.}
    \end{array}
  \right.
\end{equation}

%% file: GuA.tex
We group
the terms appearing in the equations according to the powers of $r$, and display them in increasing order in these powers.
We denote by $R[\gamma]_{AB}$ the Ricci tensor of the metric $\gamma_{AB}$.
We have: 
 \ptcheck{3VII24, GuA together with Finn against the mathematica file}
\begin{align}
	G_{uA}=&+\frac{1}{2 r^2}(n-4) D_{A}V
	\nonumber\\
	&- \frac{1}{2 r}\Biggl[\frac{(n-4) (n-3) \zhTBW_{AB} U^{B} V}{e^{2 \beta}} -  D_{A}\partial_{r}V - 2 D_{A}V \partial_{r}\beta + \zhTBW^{BC} D_{B}V \partial_{r}\zhTBW_{AC}\Biggr]
	\nonumber\\
	&+\frac{e^{-2 \beta}}{2 }\Biggl[e^{2 \beta} \Bigl(-2 D_{A}\partial_{u}\beta - 2 D_{A}U^{B} D_{B}\beta + 2 D_{A}\beta D_{B}U^{B} + 2 U^{B} (2 D_{A}\beta D_{B}\beta + D_{B}D_{A}\beta)
		\nonumber\\
		&\quad\quad\quad\quad
		 -  D_{B}D_{A}U^{B} + \zhTBW^{BC} D_{C}\partial_{u}\zhTBW_{AB}\Bigr) + (n-2) U^{B} V \partial_{r}\zhTBW_{AB}
		\nonumber\\
		&\quad\quad\quad\quad
		 + \zhTBW_{AB} \Bigl(e^{2 \beta} \zhTBW^{CD} \bigl(U^{B} (R[\gamma]_{CD} - 6 D_{C}\beta D_{D}\beta - 4 D_{D}D_{C}\beta) + D_{D}D_{C}U^{B}\bigr)
		\nonumber\\
		&\quad\quad\quad\quad
		 + 2 U^{B} V \partial_{r}\beta + (1 + n) V \partial_{r}U^{B} - 2 (n-3) U^{B} \partial_{r}V\Bigr)\Biggr]
	\nonumber\\
	&+\frac{r e^{-2 \beta}}{8}\Biggl[ -4 (n-1) \zhTBW_{BC} U^{B} D_{A}U^{C} - 4 (n-1) \zhTBW_{AC} U^{B} D_{B}U^{C} - 8 \zhTBW_{AB} U^{B} D_{C}U^{C}
	\nonumber\\
	 &\quad\quad\quad\quad
	  + 8 n \zhTBW_{AB} U^{B} D_{C}U^{C} - 4 \zhTBW^{CD} U^{B} V \partial_{r}\zhTBW_{AC} \partial_{r}\zhTBW_{BD} -  \zhTBW_{AB} \zhTBW^{CD} \zhTBW^{FG} U^{B} V \partial_{r}\zhTBW_{CF} \partial_{r}\zhTBW_{DG}
	  \nonumber\\
	  &\quad\quad\quad\quad
	   - 8 \zhTBW_{AB} V \partial_{r}\beta \partial_{r}U^{B} + 4 V \partial_{r}\zhTBW_{AB} \partial_{r}U^{B}
	  - 8 \zhTBW_{AB} U^{B} \partial_{r}\beta \partial_{r}V + 4 U^{B} \partial_{r}\zhTBW_{AB} \partial_{r}V
	  \nonumber\\
	  &\quad\quad\quad\quad
	  - 8 \zhTBW_{AB} U^{B} V \partial^2_{r} \beta + 4 U^{B} V \partial^2_{r} \zhTBW_{AB} + 4 \zhTBW_{AB} V \partial^2_{r} U^{B} - 4 \zhTBW_{AB} U^{B} \partial^2_{r} V
	  \nonumber\\
	  &\quad\quad\quad\quad
	   - 4 (n-1) U^{B} \partial_{u}\zhTBW_{AB}\Biggr]
	\nonumber\\
	&+\frac{r^2 e^{-2 \beta}}{4 }\Biggl[ -2 \zhTBW_{BC} (U^{B} \
		D_{A}\partial_{r}U^{C} + D_{A}U^{B} \
		\partial_{r}U^{C}) + 2 \zhTBW_{AC} \Bigl(2 U^{B} (2 \
		U^{C} D_{B}\partial_{r}\beta -  \
		D_{B}\partial_{r}U^{C})
		\nonumber\\
		&\quad\quad\quad\quad
		 + (2 U^{B} D_{B}\beta -  \
		D_{B}U^{B}) \partial_{r}U^{C}\Bigr) - 2 \zhTBW_{AB} \
		(\partial_{r}\partial_{u}U^{B} - 2 \
		\partial_{r}U^{B} \partial_{u}\beta) - 2 \
		\partial_{r}U^{B} \partial_{u}\zhTBW_{AB}
		\nonumber\\
		&\quad\quad\quad\quad
		+ \	2 U^{B} \Bigl(-2 U^{C} D_{C}\partial_{r}\zhTBW_{AB} -  \
		D_{C}U^{C} \partial_{r}\zhTBW_{AB} + (- D_{A}U^{C} + \
		D^{C}U_{A}) \partial_{r}\zhTBW_{BC}
		\nonumber\\
		&\quad\quad\quad\quad
		+ \zhTBW_{AB} (2 \
		D_{C}\partial_{r}U^{C} + D^{D}U^{C} \
		\partial_{r}\zhTBW_{CD} + 2 D_{C}\beta \
		\partial_{r}U^{C} + 4 \
		\partial_{r}\partial_{u}\beta) - 2 \
		\partial_{r}\partial_{u}\zhTBW_{AB}
		\nonumber\\
		&\quad\quad\quad\quad
		+ \		\zhTBW^{CD} \partial_{r}\zhTBW_{BD} \
		\partial_{u}\zhTBW_{AC} + \
		\partial_{r}\zhTBW_{AC} (- D_{B}U^{C} + D^{C}U_{B} + \
		\zhTBW^{CD} \partial_{u}\zhTBW_{BD})\Bigr)
		\nonumber\\
		&\quad\quad\quad\quad
		 + \zhTBW_{AB} \
		\zhTBW^{CD} \zhTBW^{FG} U^{B} \partial_{r}\zhTBW_{CF} \
		\partial_{u}\zhTBW_{DG}\Biggr]
		\nonumber\\
	&+\frac{r^4 e^{-4 \beta} } {4}
	(2 \zhTBW_{AC} \zhTBW_{BD} + \zhTBW_{AB} \zhTBW_{CD})  U^{B} \partial_{r}U^{C} \partial_{r}U^{D}\,,
\end{align} 

%% file: Guu.tex
%
 \ptcheck{3VII24, Guu together with Finn against the mathematica file}
\begin{align}
	G_{uu}=&
	- \frac{1}{2 r^4}(n-3) (n-1) V^2\nonumber
\\
	&+\frac{1}{2 r^3}\Bigl[e^{2 \beta} \zhTBW^{AB} \Bigl(R[\gamma]_{AB} V + 2 D_{A}\beta (- V D_{B}\beta + D_{B}V) - 2 V D_{B}D_{A}\beta + D_{B}D_{A}V\Bigr)\nonumber\\
		&\quad\quad\quad\quad
		+ (n-1) V (2 V \partial_{r}\beta -  \partial_{r}V)\Big]
		\nonumber
\\
		&- \frac{1}{8 r^2}\Bigl[8 (n-1) U^{A} V D_{A}\beta + 4 (n-7) U^{A} D_{A}V \nonumber\\
		&\quad\quad\quad\quad
		 -  V \	\bigl((4 + 8 n) D_{A}U^{A} -  \zhTBW^{AB} \zhTBW^{CD} V \
			\partial_{r}\zhTBW_{AC} \
			\partial_{r}\zhTBW_{BD} - 8 (n-1) \
			\partial_{u}\beta \bigr) - 4 (n-1) \
			\partial_{u}V\Bigr]
		\nonumber\\
		&+\frac{1}{4 r}\Bigl[\frac{2 (n-4) (n-3) \zhTBW_{AB} U^{A} U^{B} V}{e^{2 \beta}} + 4 V D_{A}\partial_{r}U^{A} + 2 V D^{B}U^{A} \partial_{r}\zhTBW_{AB} \nonumber\\
		&\quad\quad\quad\quad
		 - 4 U^{A} (D_{A}\partial_{r}V + 2 D_{A}V \partial_{r}\beta -  \zhTBW^{BC} D_{B}V \partial_{r}\zhTBW_{AC}) + 2 D_{A}V \partial_{r}U^{A}
		\nonumber\\
		&\quad\quad\quad\quad
		 - 2 D_{A}U^{A} (2 V \partial_{r}\beta + \partial_{r}V) + \zhTBW^{AB} \zhTBW^{CD} V \partial_{r}\zhTBW_{AC} \partial_{u}\zhTBW_{BD}\Bigr]
		\nonumber\\
	&- \frac{e^{-2 \beta}}{4 }\Bigl[2 (n-2) U^{A} U^{B} V \partial_{r}\zhTBW_{AB}
	\nonumber\\
		&\quad\quad\quad\quad
		 - 2 \zhTBW_{AB} U^{A} \Bigl(- e^{2 \beta} \zhTBW^{CD} \bigl(U^{B} (R[\gamma]_{CD} - 6 D_{C}\beta D_{D}\beta - 4 D_{D}D_{C}\beta) + 2 D_{D}D_{C}U^{B}\bigr)
		\nonumber\\
		&\quad\quad\quad\quad
		- 2 U^{B} V \partial_{r}\beta- 2 (1 + n) V \partial_{r}U^{B} + 2 (n-3) U^{B} \partial_{r}V\Bigr)
		\nonumber\\
		&\quad\quad\quad\quad
		 -  e^{2 \beta} \Bigl(-4 D_{A}\partial_{u}U^{A} - 2 D_{B}U^{A} (D_{A}U^{B} + D^{B}U_{A})
		\nonumber\\
		&\quad\quad\quad\quad
		 + U^{A} \bigl(8 D_{A}\partial_{u}\beta + 8 (- U^{B} D_{A}\beta + D_{A}U^{B}) D_{B}\beta - 4 \zhTBW^{BC} D_{C}\partial_{u}\zhTBW_{AB}\bigr) + 8 D_{A}U^{A} \partial_{u}\beta
		\nonumber\\
		&\quad\quad\quad\quad
		  - 4 D^{B}U^{A} \partial_{u}\zhTBW_{AB} + \zhTBW^{AB} \zhTBW^{CD} (-2 \partial_{r}\zhTBW_{AC} + \partial_{u}\zhTBW_{AC}) \partial_{u}\zhTBW_{BD}\Bigr)\Bigr]
		\nonumber
\\
	&+\frac{re^{-2 \beta}}{8}\Biggl[ 8 (n-1) \zhTBW_{BC} U^{A} U^{B} D_{A}U^{C} + \zhTBW_{AB} \biggl(-2 V \partial_{r}U^{A} \partial_{r}U^{B} + U^{A} \Bigl(8 V (2 \partial_{r}\beta \partial_{r}U^{B} -  \partial^2_{r} U^{B})
	\nonumber\\
		&\quad\quad\quad\quad
		 + U^{B} \bigl(-8 (n-1) D_{C}U^{C} + \zhTBW^{CD} \zhTBW^{FG} V \partial_{r}\zhTBW_{CF} \partial_{r}\zhTBW_{DG} + 8 \partial_{r}\beta \partial_{r}V + 8 V \partial^2_{r} \beta + 4 \partial^2_{r} V\bigr)\Bigr)\biggr)
		\nonumber\\
		&\quad\quad\quad\quad
		+ 4 U^{A} \Bigl(\zhTBW^{CD} U^{B} V \partial_{r}\zhTBW_{AC} \partial_{r}\zhTBW_{BD} -  \partial_{r}\zhTBW_{AB} (2 V \partial_{r}U^{B} + U^{B} \partial_{r}V)
		\nonumber\\
		&\quad\quad\quad\quad
		 + U^{B} \bigl(- V \partial^2_{r} \zhTBW_{AB} + (n-1) \partial_{u}\zhTBW_{AB}\bigr)\Bigr) \Biggr]
		\nonumber\\
	&- \frac{r^2 e^{-2 \beta}}{4 } U^{A} \Biggr[  \zhTBW_{BC} \Bigl(8 U^{B} (U^{C} D_{A}\partial_{r}\beta -  D_{A}\partial_{r}U^{C}) - 4 (-2 U^{B} D_{A}\beta + D_{A}U^{B}) \partial_{r}U^{C}\Bigr)
		\nonumber\\
		&\quad\quad\quad\quad
		 - 4 \bigl(\zhTBW_{AC} D_{B}U^{B} \partial_{r}U^{C} + \zhTBW_{AB} (\partial_{r}\partial_{u}U^{B} - 2 \partial_{r}U^{B} \partial_{u}\beta) + \partial_{r}U^{B} \partial_{u}\zhTBW_{AB}\bigr)
		\nonumber\\
		&\quad\quad\quad\quad
		 + U^{B} \Bigl(-4 U^{C} D_{C}\partial_{r}\zhTBW_{AB} - 2 D_{C}U^{C} \partial_{r}\zhTBW_{AB}
		\nonumber\\
		&\quad\quad\quad\quad
		 - 4 \bigl((D_{A}U^{C} -  D^{C}U_{A}) \partial_{r}\zhTBW_{BC} + \partial_{r}\partial_{u}\zhTBW_{AB} -  \zhTBW^{CD} \partial_{r}\zhTBW_{AC} \partial_{u}\zhTBW_{BD}\bigr)
		\nonumber
\\
		&\quad\quad\quad\quad
		 + \zhTBW_{AB} (4 D_{C}\partial_{r}U^{C} + 2 D^{D}U^{C} \partial_{r}\zhTBW_{CD} + 4 D_{C}\beta \partial_{r}U^{C} + 8 \partial_{r}\partial_{u}\beta + \zhTBW^{CD} \zhTBW^{FG} \partial_{r}\zhTBW_{CF} \partial_{u}\zhTBW_{DG})\Bigr)\Biggr]
	\nonumber
\\
	&- \frac{ r^4 e^{-4 \beta}}{4}(2 \zhTBW_{AC} \zhTBW_{BD} + \zhTBW_{AB} \zhTBW_{CD}) U^{A} U^{B} \partial_{r}U^{C} \partial_{r}U^{D}\,.
\end{align}

%% file: charges.tex
\section{Transport equations of $\protect\kQ{1}{}$ and $\protect\kQ{2}{}$}
\label{app6V24.1}
\subsection{$\protect\kQ{1}{}$}

From the vacuum Einstein equations we have,
\begin{equation}
         \label{22XII22.22}
  0 = \frac{r}{2(n-1)} G_{rr} = \partial_{r} \beta - \frac{r}{8(n-1)}\zhTBW^{AC}\zhTBW^{BD} (\partial_{r} \zhTBW_{AB})(\partial_r \zhTBW_{CD}),
\end{equation}
and
\begin{eqnarray}
        0
        &= &
            2r^{n-1}  G_{rA}
             \nonumber
\\
             &= &
              \partial_r \left[r^{n+1} e^{-2\beta}\zhTBW_{AB}(\partial_r U^B)\right]
            -
            2r^{2(n-1)}\partial_r \Big(\frac{1}{r^{n-1}}\spaceD_A\beta  \Big)
                 +r^{n-1}\zhTBW^{EF} \spaceD_E (\partial_r \zhTBW_{AF})
                 \,.
                 \nonumber
                 \\
                            \label{22XII22.23}
           \end{eqnarray}
Subtracting $-4 r^{n-1} \times \spaceD_A$ \eqref{22XII22.22} from \eqref{22XII22.23} gives
\begin{align}
\label{27II24.1}
    \partial_r \left[r^{n+1} e^{-2\beta}\zhTBW_{AB}(\partial_r U^B)\right] &= 2r^{2(n-1)}\partial_r \Big(\frac{1}{r^{n-1}}\spaceD_A\beta  \Big)
                 -r^{n-1}\zhTBW^{EF} \spaceD_E (\partial_r \zhTBW_{AF})
                 \nn
    \\
        & \quad
      -4 r^{n-1}  \spaceD_A\bigg[ \partial_{r} \beta - \frac{r}{8(n-1)}\zhTBW^{EC}\zhTBW^{BD} (\partial_{r} \zhTBW_{EB})(\partial_r \zhTBW_{CD})\bigg]
      \nn
      \\
      &=
      -\partial_r\big(2 r^{n-1}\spaceD_A\beta\big)
      -r^{n-1}\zhTBW^{EF} \spaceD_E (\partial_r \zhTBW_{AF})
      \nn
      \\
      &\quad
      +\frac{r^n}{2(n-1)}\spaceD_A\bigg[\zhTBW^{EC}\zhTBW^{BD} (\partial_{r} \zhTBW_{EB})(\partial_r \zhTBW_{CD})\bigg]\,.
\end{align}
Hence,
\begin{align}
    \partial_r \left[r^{n+1} e^{-2\beta}\zhTBW_{AB}(\partial_r U^B)+ 2 r^{n-1}\spaceD_A\beta\right]
      &=
      \frac{r^n}{2(n-1)}\spaceD_A\bigg[\zhTBW^{EC}\zhTBW^{BD} (\partial_{r} \zhTBW_{EB})(\partial_r \zhTBW_{CD})\bigg]
      \nn
      \\
      &\quad
      -r^{n-1}\zhTBW^{EF} \spaceD_E (\partial_r \zhTBW_{AF})\,,
      \label{23II24.31}
\end{align}
which is \eqref{23II24.21} of the main text.

\subsection{$\protect\kQ{2}{}$}
From the Einstein's equations,
 \begin{eqnarray}
         2 \Lambda  r^2
                   &=&
               r^2 e^{-2\beta} (2 G_{ur} + 2 U^A G_{rA} - V/r\, G_{rr} )
               \nonumber
\\
               & = &
                 R[\zhTBW]
                -2\zhTBW^{AB}  \Big[\spaceD_A \spaceD_B \beta
                + (\spaceD_A\beta) (\spaceD_B \beta)\Big]
                +\frac{e^{-2\beta}}{r^{2(n-2)} }\spaceD_A \Big[ \partial_r (r^{2(n-1)}U^A)\Big]
               \nonumber
\\
                &&
                 -\frac{1}{2}r^4 e^{-4\beta}\zhTBW_{AB}(\partial_r U^A)(\partial_r U^B)
                -\frac{(n-1)}{r^{n-3}} e^{-2\beta} \partial_r( r^{n-3} V)
                                 \,,
                  \label{22XII22.24}
           \end{eqnarray}
or,
\begin{align}
   & (n-1)\partial_r( r^{n-3} V)
    -2(n-1)r^{n-2}D_A U^A
    -r^{n-1}\spaceD_A \partial_r U^A
    + 2\zhTBW^{AB}e^{2\beta} r^{n-3}\spaceD_A \spaceD_B \beta
    \nn
    \\
    &
    =
    e^{2\beta}r^{n-3}\bigg[-2 \Lambda  r^2 +
                 R[\zhTBW]
                -2\zhTBW^{AB}  (\spaceD_A\beta) (\spaceD_B \beta)
                 -\frac{1}{2}r^4 e^{-4\beta}\zhTBW_{AB}(\partial_r U^A)(\partial_r U^B)\bigg]
                 \label{23II24.3}
                                 \,.
\end{align}
From \eqref{23II24.31},
\begin{align}
   &\frac{r^{n-1}}{2(n-1)}D_A \bigg[\zhTBW^{EC}\zhTBW^{BD} (\partial_{r} \zhTBW_{EB})(\partial_r \zhTBW_{CD})\bigg]
      -  r^{n-2}  \zhTBW^{EF} \spaceD_E (\partial_r \zhTBW_{AF})
      \nn
\\
    &=
      \frac{1}{r}  \partial_r \left[r^{n+1} e^{-2\beta}\zhTBW_{AB}(\partial_r U^B) + 2 r^{n-1} D_A\beta\right]
      \nn
\\
    &=
       (n+1) r^{n-1} e^{-2\beta}\zhTBW_{AB}(\partial_r U^B)
      +  r^{n} e^{-2\beta}\zhTBW_{AB}(\partial^2_r U^B)
      + \frac{1}{r} \partial_r \left[ 2 r^{n-1} D_A\beta\right]
      \nn
      \\
      &\quad
      -2r^{n} \zhTBW_{AB} (\partial_r U^B) e^{-2\beta}\partial_r\beta
       +r^{n} e^{-2\beta}\partial_r(\zhTBW_{AB})(\partial_r U^B)
      \,.
\end{align}
Multiplying by $e^{2\beta}$ and taking $D^A$ gives,
\begin{align}
    & (n+1) r^{n-1}D_B(\partial_r U^B)
      + \hspace{-1cm}
      \underbrace{ r^{n}D_B(\partial^2_r U^B)}_{r^n \partial_r (D_B\partial_r U^B) + r^n[D_B,\partial_r]\partial_r U^B}
      \hspace{-1cm}
      + \frac{1}{r} D^A\left[e^{2\beta}\partial_r \left(2 r^{n-1} D_A\beta\right) \right]
      \nn
\\
       &=
     2 r^{n}D_B[ (\partial_r U^B) \partial_r\beta]
        -r^{n}D^A \left[\partial_r(\zhTBW_{AB})(\partial_r U^B)\right]
        \nn
       \\
       &\quad
       +  D^A\left[\frac{e^{2\beta}r^{n-1}}{2(n-1)}D_A\bigg(\zhTBW^{EC}\zhTBW^{BD} (\partial_{r} \zhTBW_{EB})(\partial_r \zhTBW_{CD})\bigg)\right]
      -r^{n-2} D^A[ e^{2\beta}\zhTBW^{EF} \spaceD_E (\partial_r \zhTBW_{AF})]
      \label{23II24.2}
\end{align}
Subtracting \eqref{23II24.2} from \eqref{23II24.3} gives,
\begin{align}
    & (n-1)\partial_r( r^{n-3} V)
    -2(n-1)r^{n-2}D_A U^A
    -r^{n-1}\spaceD_A \partial_r U^A
    \nn
    \\
    &
    -(n+1) r^{n-1}D_B(\partial_r U^B)
    -r^n \partial_r (D_B\partial_r U^B)
    \nn
    \\
    &
    -r^n[D_B,\partial_r]\partial_r U^B
    - \frac{1}{r} D^A\left[e^{2\beta}\partial_r \left(2 r^{n-1} D_A\beta\right) \right]
    + 2\zhTBW^{AB}e^{2\beta} r^{n-3}\spaceD_A \spaceD_B \beta
    \nn
    \\
    &
    =
    e^{2\beta}r^{n-3}\bigg[-2 \Lambda  r^2 +
                 R[\zhTBW]
                -2\zhTBW^{AB}  (\spaceD_A\beta) (\spaceD_B \beta)
                 -\frac{1}{2}r^4 e^{-4\beta}\zhTBW_{AB}(\partial_r U^A)(\partial_r U^B)\bigg]
                 \nn
    \\
    & \quad
    - 2 r^{n}D_B [(\partial_r U^B) \partial_r\beta]
        + r^{n}D^A \left[\partial_r(\zhTBW_{AB})(\partial_r U^B)\right]
        \nn
       \\
       &\quad
       -  D^A\left[\frac{e^{2\beta}r^{n-1}}{2(n-1)}D_A\bigg(\zhTBW^{EC}\zhTBW^{BD} (\partial_{r} \zhTBW_{EB})(\partial_r \zhTBW_{CD})\bigg)\right]
      +r^{n-2} D^A[ e^{2\beta}\zhTBW^{EF} \spaceD_E (\partial_r \zhTBW_{AF})]
                                 \,,
\end{align}
where the first two lines can be rewritten to give
\begin{align}
\label{27II24.2}
    &(n-1)\partial_r
    \bigg(
    r^{n-3} V - \frac{r^{n-2}}{n-1}D^A\partial_r(r^2U_A)
    \bigg)
    - 2 r^{n-1} [D_B,\partial_r] U^B
    \nn
    \\
    &
    -r^n[D_B,\partial_r]\partial_r U^B
    -
    \frac{1}{r} D^A\left[e^{2\beta}\partial_r \left(2 r^{n-1} D_A\beta\right) \right]
    + 2\zhTBW^{AB}e^{2\beta} r^{n-3}\spaceD_A \spaceD_B \beta
    \nn
    \\
    &
    =
    e^{2\beta}r^{n-3}\bigg[-2 \Lambda  r^2 +
                 R[\zhTBW]
                -2\zhTBW^{AB}  (\spaceD_A\beta) (\spaceD_B \beta)
                 -\frac{1}{2}r^4 e^{-4\beta}\zhTBW_{AB}(\partial_r U^A)(\partial_r U^B)\bigg]
                 \nn
    \\
    & \quad
    - 2 r^{n}D_B [(\partial_r U^B) \partial_r\beta]
        + r^{n}D^A \left[(\partial_r\zhTBW_{AB})(\partial_r U^B)\right]
        \nn
       \\
       &\quad
       -  D^A\left[\frac{e^{2\beta}r^{n-1}}{2(n-1)}D_A\bigg(\zhTBW^{EC}\zhTBW^{BD} (\partial_{r} \zhTBW_{EB})(\partial_r \zhTBW_{CD})\bigg)\right]
      +r^{n-2} D^A[ e^{2\beta}\zhTBW^{EF} \spaceD_E (\partial_r \zhTBW_{AF})]
                                 \,.
\end{align}
We rewrite the second term in the second line of \eqref{27II24.2} as
\begin{align}
    \frac{1}{r} D^A\left[e^{2\beta}\partial_r \left(2 r^{n-1} D_A\beta\right) \right]
    &=
    \partial_r\bigg(2 r^{n-2}  e^{2\beta} \Delta \beta \bigg)
    +\partial_r( 2 r^{n-2}  D^A(e^{2\beta}) D_A \beta) 
    \nn
    \\
    &\quad + 2 r^{n-2}   [D^A,\partial_r](e^{2\beta} D_A \beta) - 2 r^{n-1} 
 D^A[\partial_r(e^{2\beta}/r)D_A\beta]
    \,,
\end{align}
which can be substituted back into \eqref{27II24.2} to give
\begin{align}
\label{27II24.2b}
    &(n-1)\partial_r
    \bigg(
    \underbrace{r^{n-3} V - \frac{r^{n-2}}{n-1}D^A\partial_r(r^2U_A) - \frac{2 r^{n-2}}{n-1}  e^{2\beta} \Delta \beta}_{=:  \chi}
    \bigg)
    \nn
    \\
    &=
    2 r^{n-1} [D_B,\partial_r] U^B + \partial_r( 2 r^{n-2}  D^A(e^{2\beta}) D_A \beta) 
    \nn
    \\
    &\quad + 2 r^{n-2}   [D^A,\partial_r](e^{2\beta} D_A \beta) - 2 r^{n-1} 
 D^A[\partial_r(e^{2\beta}/r)D_A\beta]
    \nn
    \\
    &\quad
    +r^n[D_B,\partial_r]\partial_r U^B
    - 2\zhTBW^{AB}e^{2\beta} r^{n-3}\spaceD_A \spaceD_B \beta
    \nn
    \\
    &\quad
    +
    e^{2\beta}r^{n-3}\bigg[-2 \Lambda  r^2 +
                 R[\zhTBW]
                -2\zhTBW^{AB}  (\spaceD_A\beta) (\spaceD_B \beta)
                 -\frac{1}{2}r^4 e^{-4\beta}\zhTBW_{AB}(\partial_r U^A)(\partial_r U^B)\bigg]
                 \nn
    \\
    & \quad
    - 2 r^{n}D_B [(\partial_r U^B) \partial_r\beta]
        + r^{n}D^A \left[(\partial_r\zhTBW_{AB})(\partial_r U^B)\right]
        \nn
       \\
       &\quad
       -  D^A\left[\frac{e^{2\beta}r^{n-1}}{2(n-1)}D_A\bigg(\zhTBW^{EC}\zhTBW^{BD} (\partial_{r} \zhTBW_{EB})(\partial_r \zhTBW_{CD})\bigg)\right]
      +r^{n-2} D^A[ e^{2\beta}\zhTBW^{EF} \spaceD_E (\partial_r \zhTBW_{AF})]
                                 \,.
\end{align}